\documentclass{aa}  

\def\msun{\mbox{$ M_\odot\,$}}
\def\lsun{\mbox{$\rm L_\odot\,$}}

\def\mum{\mbox{$\rm \,\mu m\,$}}
\def\herschel{$\it{Herschel}$\,}
\def\planck{$\emph{Planck}$\,}
\def\deg{\hbox{$^\circ$}}
\def\arcmin{\hbox{$^\prime$}}
\def\arcsec{\hbox{$^{\prime\prime}$}}

\usepackage{natbib}
\usepackage{graphicx}
\usepackage{txfonts}
\begin{document} 

 \title{Filament coalescence and hub structure in Mon\,R2\thanks{Based on 
observations obtained with \herschel programme HOBYS} \thanks{Based on observations obtained with \planck \, (http://www.esa.int/Planck), an ESA science mission with instruments and contributions directly funded by ESA Member States, NASA, and Canada.} }

\titlerunning{Filament coalescence and hub structure in Mon\,R2}

   \subtitle{Implications for massive star and cluster formation}

   \author{M. S. N. Kumar\inst{1}, D. Arzoumanian\inst{2,1}, A. Men'shchikov\inst{3}, P. Palmeirim\inst{1}, M. Matsumura\inst{4}
          \and
          S. Inutsuka\inst{5} }

   \institute{Instituto de Astrof\'{i}sica e Ci\^{e}ncias do Espa\c{c}o,
     Universidade do Porto, CAUP, Rua das Estrelas, PT4150-762 Porto, Portugal
     \and Aix Marseille University, CNRS, CNES, LAM - 38 rue Frederic Joliot-Curie, 13388 Marseille, France
     \and AIM, IRFU, CEA, CNRS, Universit\'{e} Paris-Saclay, Universit\'{e} Paris Diderot, Sorbonne Paris Cit\'{e}, F-91191 Gif-sur-Yvette, France     \and Faculty of Education \& Center for Educational Development and Support, Kagawa University, Saiwai-cho 1-1, Takamatsu 760-8522, Japan
     \and Department of Physics, Nagoya University, Furo-cho, Chikusa-ku, 
Nagoya, Aichi 464-8602, Japan
     \\ \email{nanda@astro.up.pt}
            }
\authorrunning{Kumar, Arzoumanian, Men'shchikov, Palmeirim, Matsumura \& Inutsuka}

 
  \abstract {There is growing evidence of the role of hub-filament systems (HFS) in the formation of stars from low to high masses. As of today, however, the detailed structures of these systems are still not well described. Here we study the Mon\,R2 star-forming region, which has a rich network of filaments joining in a star cluster forming hub, and aim to understand the hub structure and to examine the mass fraction residing in the hub and in the filaments, which is a key factor that influences massive 
star formation. We conducted a multi-scale, multi-component analysis of the \herschel column density maps (resolution of 18.2\arcsec\, or $\sim$0.07\,pc at 830\,pc) of the region using a newly developed algorithm \textsl{getsf} to identify the structural components, namely, extended cloud, filaments, and sources. We find that cascades of lower column density filaments coalesce to form higher-density filaments eventually merging inside 
the hub (0.8\,pc radius). As opposed to the previous view of the hub as a 
massive clump with $\sim$1\,pc radius, we find it to be a network
of short high-density filaments. We analyse the orientations and mass per 
unit length (M/L) of the filaments as a function of distance from the hub 
centre. The filaments are radially aligned towards the centre of the hub. 
The total mass reservoir in the Mon\,R2 HFS (5\,pc $\times$ 5\,pc) is split between filaments (54\%), an extended cloud (37\%), and sources (9\%). The 
M/L of filaments increases from $\sim$ 10\msun/pc at 1.5\,pc from the hub to $\sim$ 100 \msun/pc at its centre, while the number of filaments per annulus of 0.2\,pc width decreases from 20 to two in the same range. The observed radial column density structure of the HFS (filament component only) displays a power-law dependence of $N_{\mathrm{H}_2} \propto r^{-2.17}$ 
up to a radius of $\sim$2.5\,pc from the central hub, resembling a global 
collapse of the HFS. We present a scenario where the HFS can be supported by magnetic fields which interact, merge, and reorganise themselves as the filaments coalesce. We plotted the plane-of-the-sky magnetic field line geometry using archival \planck data to support our scenario. In the new view of the hub as a network of high-density filaments, we suggest that only the stars located in the network can benefit from the longitudinal flows of gas to become massive, which may explain the reason for the formation of many low-mass stars in cluster centres. 
We show the correlation of massive stars in the region to the intertwined 
network-like hub, based on which we updated the implications of the filaments to clusters (F2C) model for massive star formation.}


   \maketitle
%

\section{Introduction}

The interstellar medium (ISM) is observed to be filamentary both
in its atomic \citep{Heiles79} and molecular \citep{SchneiderElmegreen79} 
forms. This view is firmly pronounced by
data obtained with modern observational facilities with ever-increasing angular resolution and sensitivities \citep{Andrepp6, Hacar2018, Arzoumanian2019}. A hierarchy of
filamentary structures have been named to describe observational data:
galactic spines \citep{Goodman14}, giant filaments \citep[e.g.][]{Ragan14,Zucker2018}, interstellar filaments \citep[e.g.][]{Arzoumanian2019},
fibres \citep[e.g.][]{TafallaHacar15,Hacar17}, and striations and strands \citep[e.g.][]{cox16,Bonne20}. This hierarchy is distinguished by the
physical size and the total mass held within them. 

In the nearby ($<$500pc) star-forming regions, filamentary structures span a range of $\sim$3 orders of magnitude in surface densities \citep{Arzoumanian2019}. The resulting filament mass functions are proposed to be the origin of dense core- and stellar-mass functions \citep{Andre19}. Velocity coherent filaments, also known as fibres \citep{Hacar17,Hacar2018} characterised in NGC1333 and the Orion integral shaped filament (ISF), indicate that the fibre density increases with the total mass per unit length of filaments.

\begin{figure*}[ht]
\centering
\includegraphics[width=\linewidth]{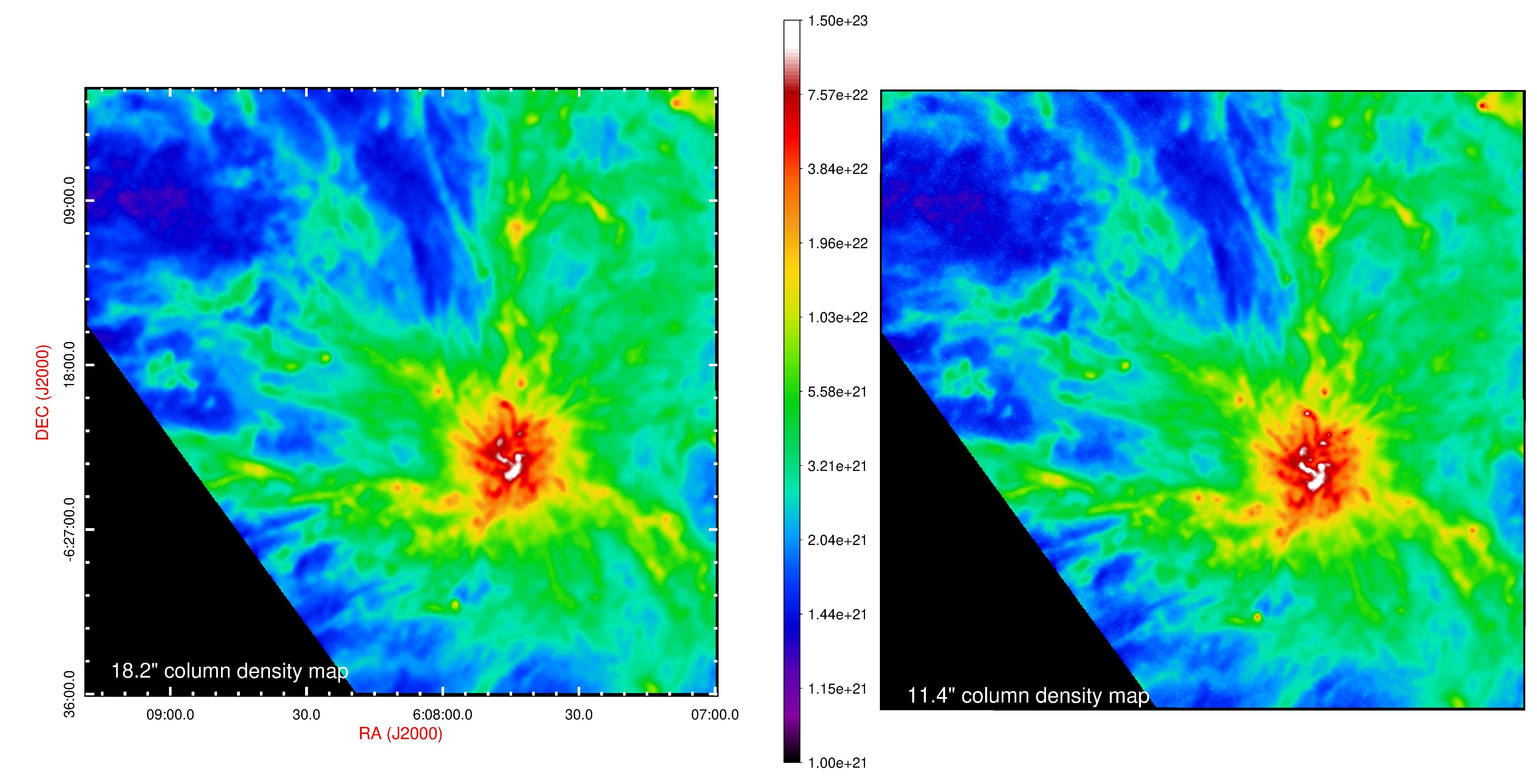}
\caption{Column density maps of the Mon\,R2 star-forming region at two angular resolutions, produced using \textsl{getsf}. The vertical colour bar is in units of cm$^{-2}$. At a distance of 830\,pc, the 11.4\arcsec\ 
resolution corresponds to a beam of 0.046\,pc.}
\label{fig:SD}
\end{figure*}

Velocity coherent dense cores \citep{Barranco1998} have a radius of $\sim$0.1\,pc, a scale at which the structure decouples itself from the larger-scale turbulent flows \citep{Goodman98} and transits to a coherent sonic entity. Interestingly, the constant widths of filaments of $\sim$0.1\,pc, 
as inferred from \herschel observations of nearby star-forming regions \citep{Andrepp6,Arzoumanian11,Arzoumanian2019}, represent a wide range in mass per unit length (M/L) of filaments, most remarkably that of a 1000\msun\,pc$^{-1}$ filament in NGC6334 \citep{Andre16}. This coincidence makes it tempting to view filaments as cylindrical equivalents of coherent dense cores.  \citet{Hacar16,Hacar2018} suggest that velocity coherent fibres 
may represent the basic structure in star-forming clouds, the fibre spatial density distinguishing low- and high-mass star-forming regions. However, there is also no guarantee that structures traced in PPV space correspond to real structures in PPP space \citep[e.g.][]{clarke18}. Given that there is no well-defined boundary that separates velocity coherence and that fibres or fibre bundles are identified as filaments, these structures, in general, may represent a scale at which they decouple themselves from the 
cloud-scale turbulence. 

Junctions of filaments on the other hand are called hubs, which are high column density low aspect ratio objects \citep{Myers2009, schneider2012,Peretto14,Chen19,Kumar2020}. Massive star formation itself appears to take 
place only in the hubs \citep{Kumar2020, schneider2012}, which are found to be fed by longitudinal flows along filaments \citep{Peretto14, Williams18}. \citet{Kumar2020} note a hierarchy of hub-filament systems (HFS) which they speculate to be a consequence of the hierarchical range in filamentary structures from fibres to galactic spines. The identification of core- and clump-hubs in NGC6334 \citep{Arzoumanian2021} is a manifestation of that hierarchy in HFS. 
 
The property of hubs as density-enhanced objects \citep{Kumar2020} and the finding that fibre dimensions are self-regulated depending on their intrinsic gas density \citep{Hacar2018} hints at a mechanism that may be at work universally in building the density structure of star-forming regions. If high M/L filaments are dense bundles of fibres and density enhanced 
hubs are junctions of filaments, it is necessary to understand the relation between the two and 
which type of junction leads to what structure. Next, it appears that all massive star formation in the Milky-Way plane covered by the \herschel Hi-GAL survey \citep{Molinari2010} takes place in density amplified hubs \citep{Kumar2020}, and the hubs are replenished by material flow from filaments exterior to the hub and constituting the HFS. While the filaments to clusters (F2C) model of \citet{Kumar2020} assumes that this flow of matter originates in longitudinal flows \citep{Peretto14,Chen19}, a global flow of accelerated matter is reported in the Orion ISF by \citet{Hacar17b} who argue for a global collapse. If so, it is necessary to estimate the fraction of material held in the filaments (outside the hub) of any HFS and to asess if it is significant enough to aid the massive star formation or enhance cluster formation in the hubs.

To answer these two questions, we chose to study the Mon\,R2 star-forming 
region which has a rich network of filaments merging into a hub \citep{TrevinoMorales2019}, a nearly face-on geometry, and the region as a whole displays a column-density probability distribution function N-PDF \citep{Pokhrel16} that is representative of any cluster-forming cloud. Our analysis is based on a multi-scale multi-component analysis, which uses a new algorithm \textsl{getsf} \citep{sasha20} to separate the elongated filaments from extended background emission and roundish sources. This is crucial 
because we are interested in analysing the properties of the filaments as 
independent entities, excluding contributions from the embedded sources and the surrounding diffuse cloud.

This paper is organised as follows. In Sec.\,2 we present the observational data and column density maps used for the study. Analysis of the data is described in Sec.\,3, which involves spatially decomposing the filament, extended cloud, and sources (sec.\,3.1), classifying the hub and filament components of the HFS, computing their mass fractions (sec.\,3.2), and studying the filament lengths, angles, and mass-per-unit-length as a function of its distance to the hub centre (sec.\,3.3). Interpretations based on this analysis are deliberated in Sec.\,4, which include coalescence of molecular filaments (sec.\,4.1), the HFS mimicking global collapse (sec.\,4.2), the hub-filament mass fraction (sec.\,4.3), the structure of the hub and its implications for the formation of massive stars 
(sec.\,4.4), and the role of magnetic fields as necessary support against global collapse (sec.\,4.5). A discussion reflecting on the meaning of our interpretations to the larger picture can be found in Sec.\,5.

\section{Observational data}

Among the nearby (< 1\,kpc) star-forming regions, Mon\,R2 ($d$ = 830\,pc) is a hub-filament system with a miniature spiral galaxy appearance \citep{TrevinoMorales2019}, where the dense central hub is forming a cluster 
of stars including the B-type star at the centre of the Mon\,R2 region (Fig.1). This large network of filaments is shown to be constrained in a flattened sheet-like space, based on an analysis of the velocity information from the observations of molecular emission lines \citep{TrevinoMorales2019}. The sheet has a low inclination angle of $\sim$30\deg\, with the plane of the sky giving its spiral galaxy appearance. In this study, we have produced 18.2 \& 11.4\arcsec\, high-resolution surface density maps and separated the structural components of the maps using \textsl{getsf} \citep{sasha20} to find filaments, sources, and the extended non-filamentary 
emission (hereafter diffuse cloud). This allowed us to derive an intrinsic surface density structure and the mass of the individual filaments, free of the contributions from the unrelated structural components. Our analysis is 
different from the previous studies that examined the radial profiles of the overall density \citep{Didelon2015} or simple estimates of the diffuse component based on measurements in regions void of filaments \citep{TrevinoMorales2019}. Here the dense and diffuse cloud components are separated from the sources by decomposition of the observed data, allowing us to examine the behaviour of each component and its impact on the physics of star formation.

The \herschel data of Mon\,R2 taken as part of the HOBYS programme \citep{motte2010} provide imaging in five far-infrared bands at 70, 160, 250, 350, 
and 500\mum. Early results on Mon\,R2 using HOBYS data are described by \citet{Didelon2015} and \citet{Rayner2017}. These images were used to compute the surface density at the standard resolution of 36\arcsec\, and a higher angular resolution of 18.2\arcsec\,\citep{Palmeirim13}, corresponding to 
the resolution of the 250\mum\, image. Using the \textsl{hires} method described by \citet{sasha20}, a 11.4\arcsec\, resolution map (corresponding 
to the 160\mum\, beam) was produced. These maps are displayed in Fig.\,1. 
Because the 160\mum\, might also contain contributions from the hotter dust emission, we tested the fidelity of the surface density map at 11.4\arcsec\, resolution by smoothing it to 18.2\arcsec\, and 36\arcsec\, and comparing with the 18.2\arcsec\, high-resolution and 36\arcsec\, standard resolution maps. While the total surface density remained conserved, the maps at different angular resolutions were locally comparable with differences of 10-20\%. The column densities, derived from the spectral energy density (SED) fitting of the \herschel data assuming a uniform dust temperature along the line of sight, may be affected by possible temperature gradients along the line of 
sight, especially towards star-forming protostellar sources \citep{Roy2014,sasha20} and \ion{H}{ii} regions. If higher temperature dust is present 
along the line of sight, the best-fitted column densities are only underestimated \citep{Anderson2012}, especially towards the densest filaments and hubs embedded in a warmer cloud envelope. The structures traced by both the 11.4\arcsec\, and 18.2\arcsec\, hi-resolution maps shown in Fig.\,1 
are almost identical. We traced the structures using the 11.4\arcsec\, map 
and obtained the column density measurements from the 18.2\arcsec\, map.\\

\begin{figure*}[ht]
\centering
\includegraphics[width=\linewidth]{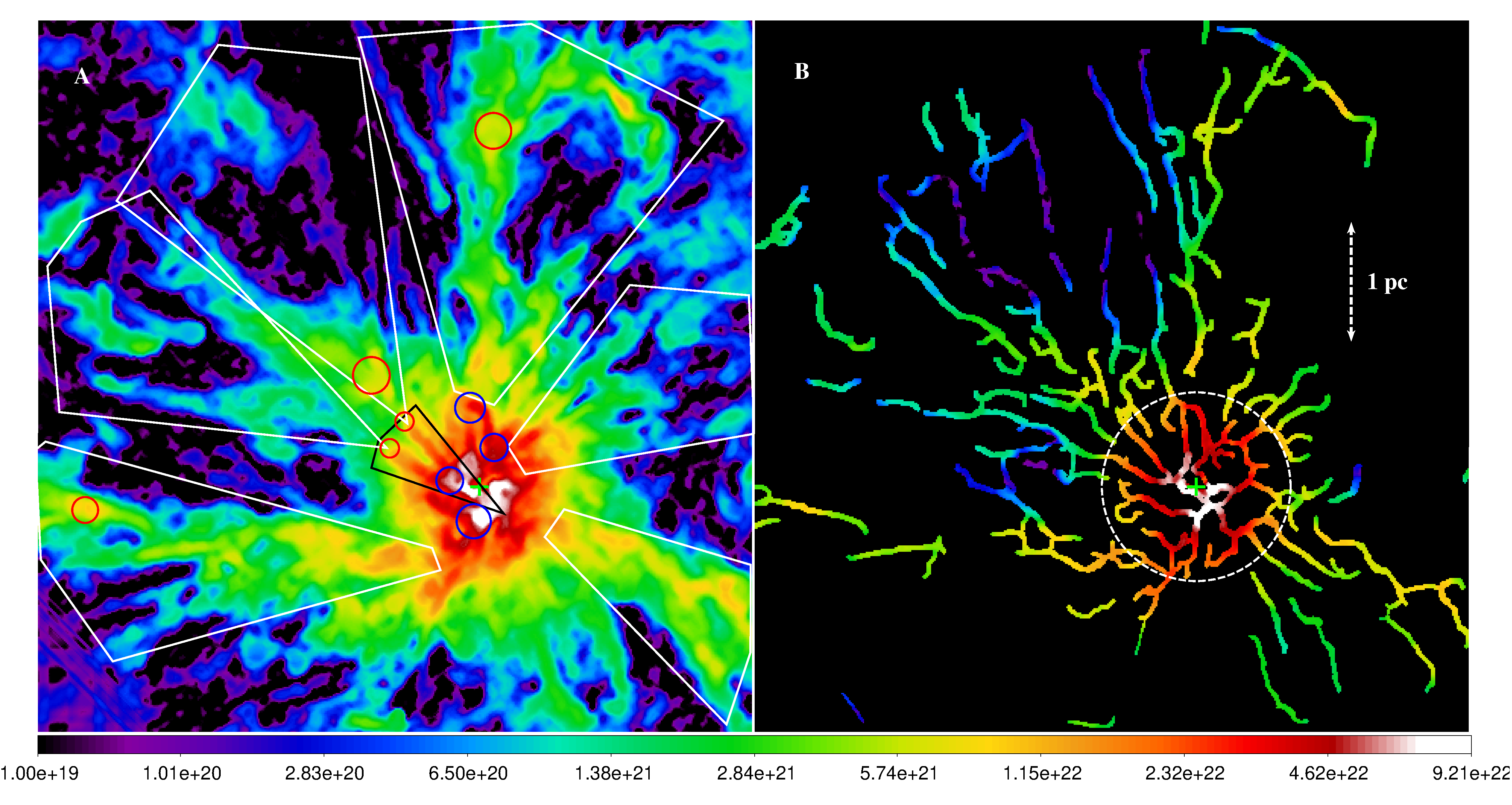}
\caption{ Coalescence of filaments: a) Colour map showing the column density of the filament component that is the sum of filaments detected over scales from 16\arcsec\, up to 258\arcsec\,. Groups of coalescing filaments (see Sec.\,4.1) are enclosed by white line boundaries. The black cone encloses an example of a second step in the coalescence. The red and blue circles (colours used for convenient visual contrast) identify some of the clear junctions of filaments within the coalescing groups where the density is visibly enhanced 
compared to the incoming structures. b) Individual filament crests identified by \textsl{getsf} are used to mask the image in panel a. In both panels, the cross symbol at the centre (black and green in the left 
and right panels, respectively) indicates the position of the hub centre and radial origin. }
\label{fig:skel}
\end{figure*}

\section{Analysis}
\subsection{Filaments, sources, and extended cloud}
\subsubsection{Spatial decomposition}

The filaments, compact sources, and the large-scale diffuse background were 
decomposed on the surface density maps using the \textsl{getsf} algorithm 
\citep{sasha20,sasha21}. The method was developed for extracting filaments and sources in far-infrared to sub-millimetre observations with the \herschel Space Telescope, nevertheless, it applies to any data set. A unique feature of \textsl{getsf} is the careful separation of the structural components based on their shapes. In this method, the spatial decomposition of the observed image is conducted in a range of scales to identify the 
structural components and separate them. This allows one to disentangle features over multiple scales and to separate the sources from filaments and backgrounds. The main processing steps of \textsl{getsf} are the spatial decomposition of the observed images, separation of the structural components, flattening of the detection images of the source and filament components, the combination of the detection images from different wavebands, detection of sources and filaments, and their measurements. 
We applied \textsl{getsf} to the surface density maps of Mon\,R2 using the 11.4\arcsec\, resolution map to detect the structural components and obtain the measurements on the 18.2\arcsec\, map. This way we optimally utilised the angular resolution in the data to accurately trace the morphological characteristics while retaining the column density measurements from the 18.2\arcsec\, map.

\subsubsection{Filament component map}
The output from the \textsl{getsf} decomposition includes three component 
images, namely filaments, extended emission background, and compact sources, a sum of which corresponds to the original surface density map. All of these 
images are displayed in the appendix Fig.\,A.1. The image of filaments is 
a reconstruction of filaments detected on scales ranging from 16 to 258\arcsec, the upper limit defined after an initial examination of the features in the input image. The skeletons tracing crests of the detected filaments on different angular scales, as well as catalogues containing lengths, widths, surface densities, and linear densities, are produced by \textsl{getsf} \citep{sasha20}. 

 From Figs.\,1 \& 2, it is evident that most filamentary structures have similar widths and there are not many structures with largely discrepant widths. This is because the MonR2 is at a distance of 830\,pc and the angular resolution of 11--18\arcsec\, cannot resolve physical scales smaller 
than 0.1\,pc. Even though identification was made over the scales between 
16 to 258\arcsec, most of the filaments and sources are detected on scales below 50\arcsec.  A given filament is often traced on more than one scale depending on its width. Almost all filaments in the region displayed in Fig.\,1 are exhaustively traced by the scales up to 46\arcsec\, corresponding to the physical size of $\sim$0.18\,pc.  On scales larger than 46\arcsec\,, the features mostly represent the lower-column-density and larger-scale halo structures surrounding groups of multiple filaments. To analyse the filaments, we used a set of unique skeletons that were identified up to an angular scale of 46\arcsec. A mask was created from those unique skeletons and applied to the filament component image to examine the crest column densities and their variation in the entire image, as shown in Fig.\,2b. The column densities displayed in Fig.\,2a represent only the contribution from the filaments (dense gas, $A_{V}>8$\,mag), after the removal of sources and the variable diffuse background.

\begin{figure}[ht]
\centering
\includegraphics[width=\linewidth]{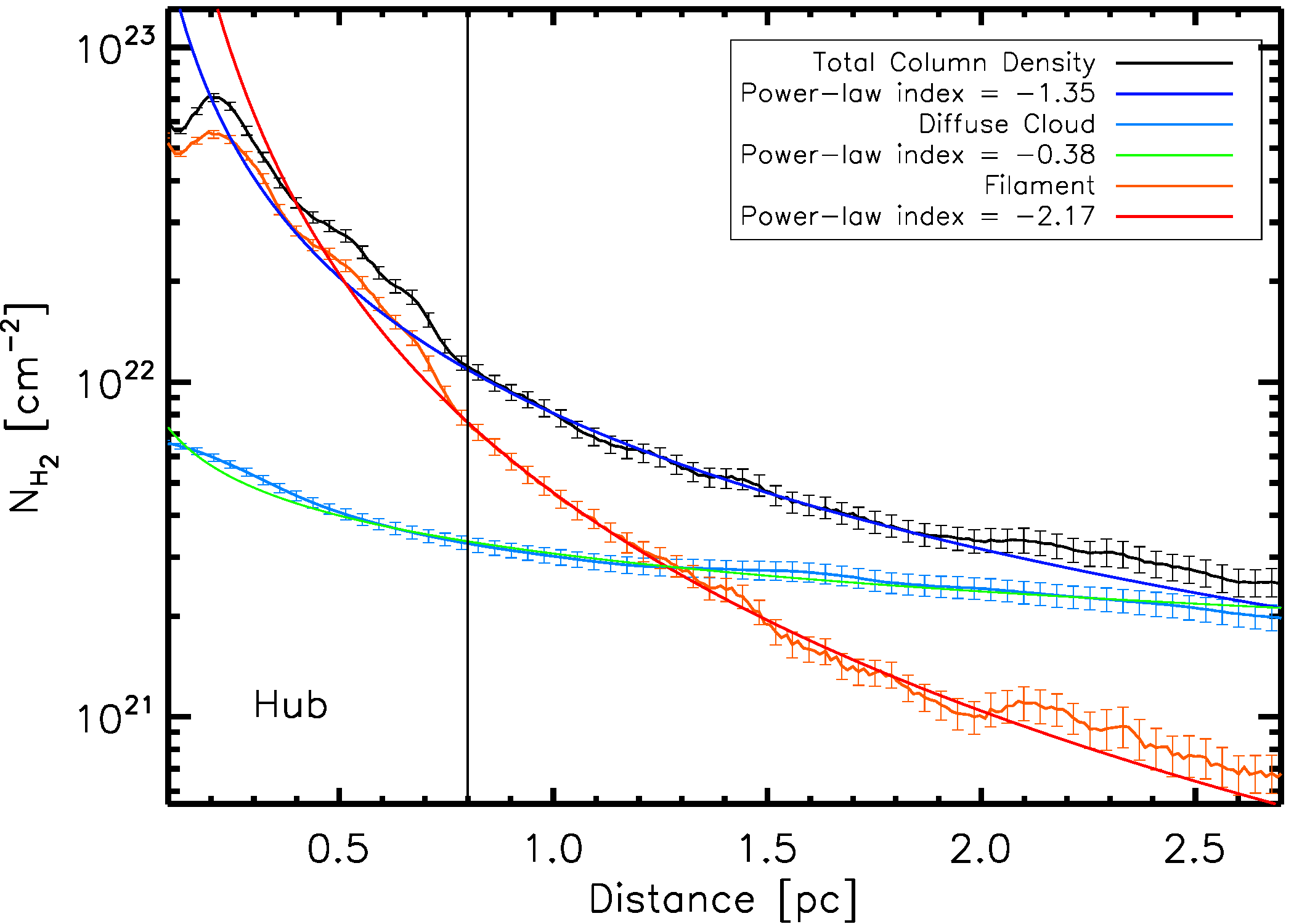}
\caption{Radial profiles of the azimuthally averaged column density centred on the Mon\,R2 hub, as distributed in its structural components. The column density radial profile of the total map is compared with separated filaments and diffuse cloud component maps. The fitted curves and their power-law indices are listed for each component. The fits ware performed for regions beyond 1\,pc, excluding the hub. Error bars represent the standard deviation of the non-zero values in each pixel annulus.}
\label{fig:radprof}
\end{figure}

\begin{figure}[ht]
\centering
\includegraphics[width=\linewidth]{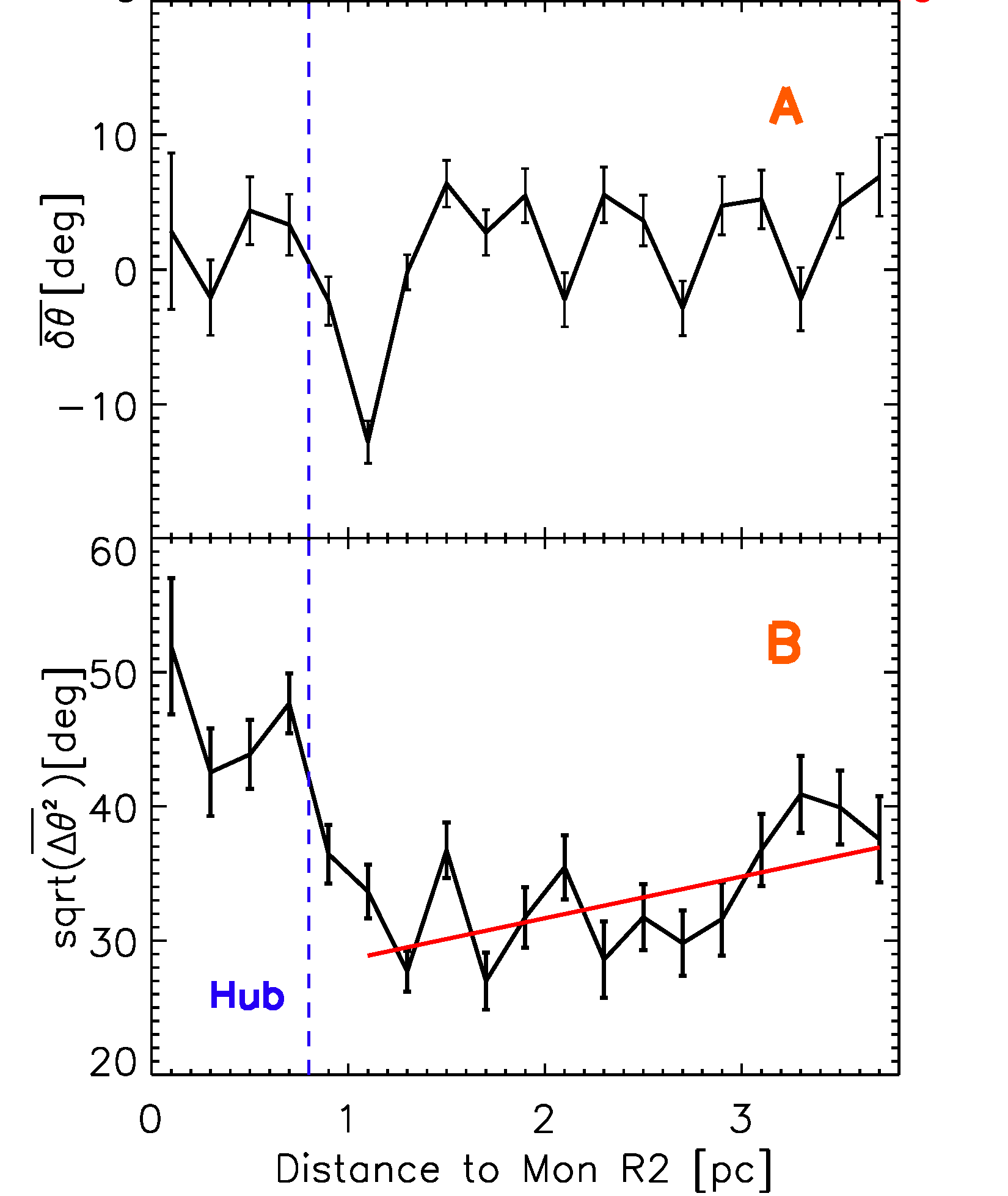}
\caption{Filament angle analysis. a) Average filament orientation with respect to the purely radial direction as a function of distance to the centre of the hub shows that 
they are radially aligned. b) Deviation of the filament angle from the purely radial direction to Mon\,R2 is an increasing function of distance to 
Mon\,R2, suggesting influence from the hub gravitational potential in aligning the filaments.}
\label{fig:angle}
\end{figure}

\begin{figure}[ht]
\centering
\includegraphics[width=\linewidth]{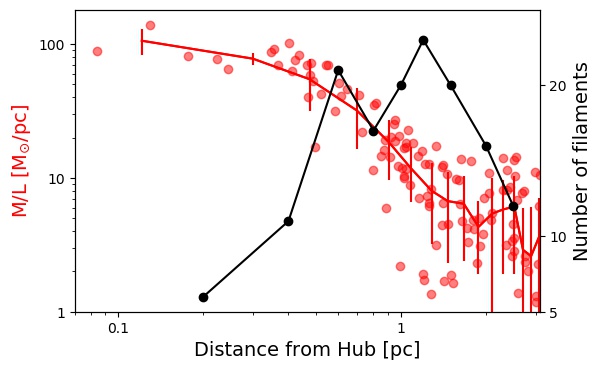}
\caption{Radial variation of filament M/L (red) and the number of filaments crossing annular circles (black). Red dots show the actual data points for all skeletons longer than 0.4\,pc. The red line represents the average M/L in each 0.2\,pc bin and the error bars are their standard deviations. The black line shows the number of filaments  that intersect the circumference of concentric circles drawn around the HFS as displayed in Fig.\,A.2.}
\label{fig:linemass}
\end{figure}
\begin{figure}[ht]
\centering
\includegraphics[width=\linewidth]{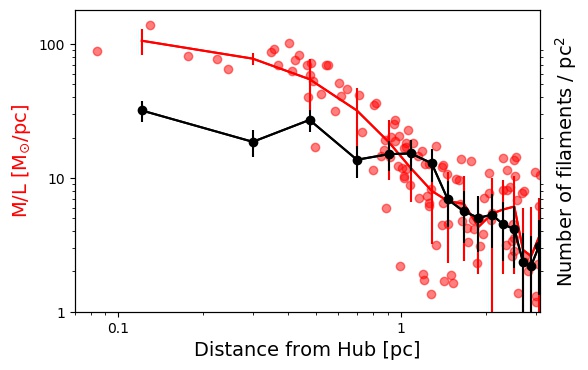}
\caption{Different version of Fig.\,5 where the number of filaments per 
unit area (N) is compared with the M/L. Readers should notice that while both the M/L and N increase with decreasing distance from the hub centre from 3\,pc to 1\,pc, in the region within 1\,pc, N flattens out while M/L steeply continues to increase with decreasing distance before flattening out in the central 0.2-0.3\,pc.}
\label{fig:N-ML}
\end{figure}

\subsection{Characterising the Mon\,R2 hub-filament system}

Given the circular symmetry of the Mon\,R2, we have analysed its properties in radial coordinates, centred on the location of the B star in Mon\,R2, which is also the centre of the spiral-like pattern of the column density map. Throughout the subsequent analysis, the location of this radial origin is considered to be $\alpha_0$ = 06:07:45.6 $\delta_0$ = $-$6:23:00.

\subsubsection{The hub and its relation to the stellar cluster}
We first studied the azimuthally averaged radial surface density of the structural components. The extended background is assumed to represent the 'diffuse cloud' from now on. The column density radial profile of the non-decomposed original image as well as the decomposed filaments- and diffuse-components in a 5\,pc $\times$ 5\,pc projected region centred on the radial origin were produced (Fig.\,3). From the radial origin ($\alpha_0$, $\delta_0$), moving outwards, a one-pixel width azimuthal average was computed up to a projected radial distance of 2.5\,pc. The standard deviation of the non-zero pixels in 
the azimuthal average was used as the error at each radial point. The curves were fitted with power-law functions for the region with a radius between 0.8\,pc and 2\,pc, allowing the power-law exponent to be freely adjusted by the fitting programme. Also, we note that the measurements for filaments here involve subtracting the diffuse component, hence, the low values of N$_{H_2} < $10$^{21} cm^{-2}$ beyond 2.0\,pc. The fitted functions are also plotted in Fig.\,3. As evident from this figure, the region interior to 0.8\,pc, defined as the hub, shows a steeper change in the column densities, whereas the region beyond the hub varies smoothly. Therefore, we define the radius of the hub R$_{\rm hub}$=0.8\,pc. We define the inner 0.4\,pc as the hub-core for the analysis here.

The Mon\,R2 cluster identified by \citet{carpenter1997} is centred on 06:07:46.2 $-$6:22:51.6, with a full-width-half-maximum size scale of 0.38$\pm$0.03\,pc and an estimated central stellar volume density of $\sim$9000$\pm$1000 stars pc$^{-3}$. Even though the cluster extent is traced up to 2\,pc, the stellar 
surface density effectively drops to $<$ 50 stars pc$^{-2}$ at a radius of 
0.5\,pc \citep[see Fig.\,7 in][]{carpenter1997}. This radius is similar to the radius of the hub-core above, suggesting that much of the observed star formation resides within the hub-core. This region corresponds to the area represented by the highest-column-density filaments appearing as white skeletons in Fig.\,2b.

\subsubsection{Mass fraction of the structural components}

\begin{table}[ht]
\centering
\caption{\label{tab:example}Non-cumulative mass of the different components (filaments, diffuse cloud, and compact sources) as a function of radius.}
\begin{tabular}{|l|l|l|l|l|}
\hline
Radius (pc)& 0.4 & 0.4--0.8 & 0.8--1.6 & 1.6--2.5 \\
\hline
& \multicolumn{4}{c}{Mass \msun} \\
\hline
Filament & 837 & 761 & 632 & 345 \\
Diffuse cloud & 108 & 198 & 585 & 855 \\
Sources & 164 & 115 & 103 & 71 \\
Total & 1109 & 1074 & 1320 & 1271 \\
\hline
4774\msun=100\% & \multicolumn{4}{c}{Mass fraction \% } \\
\hline
Filament (53.8) & 17.5 & 15.9 & 13.2 & 7.2 \\
Diffuse cloud (36.7) & 2.3 & 4.2 & 12.2 & 18.0  \\
Sources (9.5) & 3.4 & 2.4 & 2.2 & 1.5 \\
\hline
\end{tabular}
\end{table}

Table\,1 lists the computed mass of the individual structural components 
distributed in four annular regions centred on the radial origin, which is also the hub centre. The radii of 0.4 and 0.8\,pc represent the central 
core of the hub and the hub.  The mass reservoir for all structural components listed in Table\,1 were obtained by directly converting the observed 
column density (N$_{H_2}$) for each of these components and assuming a distance of 830\,pc to Mon\,R2  by using the formula Mass = N$_{H_2} \times$ area$\times\, \mu_{H_2} \times\, m_H $, where $\mu_{H_2}$ is the mean 
molecular weight per H$_2$ taken as 2.8 and m$_H$ is the atomic weight of 
hydrogen. These values are generally in agreement with previous estimates 
\citep{TrevinoMorales2019}  if we take the differences in the analysed area of the image into account. However, here, the diffuse cloud and source masses are more accurately derived owing to the proper decomposition of these components.  We note that we have restricted this table to a radial distance of 2.5\,pc (also true for Figs.\,4 and 5), because of the symmetrical field-of-view centred on Mon\,R2 which is available from the data used. 
The cumulative mass within a circular region of radius 2.5\,pc combined from all components is 4774\msun, which was considered to be 100\% while computing the listed mass fraction in Table.\,1. It can be seen that about 54\% of the total mass reservoir resides in the filaments and nearly 40\% in the diffuse cloud. The diffuse cloud mass is often found to be higher as estimated by the CO observations (see Sect. 4.3). Considering only the filaments' component, the central hub region within the radius of 0.8\,pc holds $\sim$ 1598 \msun, while $\sim$977 \msun is held in the filaments located 
in the annular region of 0.8-2.5\,pc. This indicates a filament:hub mass ratio of 0.6:1. However, the filaments extend further beyond to a radius 
of more than 4\,pc from the centre, so the filaments outside the hub should hold a mass much larger than 977\msun\, if not twice that value because 
the column density of the filaments decreases as $\sim$r$^{-2.17}$ (see Fig.\,3) away from the hub. Therefore, it is reasonable to assume that the 
mass residing in the hub area is roughly similar to the mass residing in the filaments exterior to the hub region. 

\subsection{Filament lengths, angles, and line masses}

We analyse the orientation of the filaments to examine in detail the circularly symmetric pattern of the Mon\,R2 region. The radial origin defined 
before is also the centre of the spiral-like pattern of the column-density map. To assess, how the filaments are oriented around Mon\,R2, we divided the region around Mon\,R2 radially into many concentric annuli with widths of $dr$=0.2\,pc. In each annulus, we calculated the average deviation $\overline{\delta\theta}$ of the filament orientation $\theta$ from the purely radial direction from the hub centre, as well as  $\Delta\theta = \delta\theta -\overline{\delta\theta}$ and average $\overline{\Delta \theta}$. Plotting $\overline{\delta\theta}$ and $\sqrt{\overline{{\Delta\theta}^2}}$ as a function of the radial distance of the annuli, we note that $\overline{\delta\theta}$ remains close to zero up to $\sim$3\,pc, suggesting that the filaments are oriented radially towards Mon\,R2 (Fig.\,4a). The $\sqrt{\overline{{\Delta\theta}^2}}$ is a slowly increasing function of radius beyond the 'hub' (Fig.\,4b), suggesting that the filaments deviate from the purely radial direction away from the hub. 

Next, the radial variation of the filament mass per unit length (M/L) and 
the number density is investigated. For every skeleton, the average of the radial distances of all skeleton pixels was taken as the filament distance from the hub centre and origin. The crest column densities for each skeleton were measured and averaged using the masked skeleton map shown in Fig.\,2b. The total mass of each filament was computed by assuming a 0.1\,pc width \citep{Arzoumanian11} and 830\,pc distance to Mon\,R2. The total 
mass was divided by skeleton lengths to obtain the M/L of the filaments. An average value computed in 0.2\,pc bins were used to produce the red curve in Figs. 5 \& 6, and the error bars are the standard deviation of the average.  Similarly, the total number of filaments in 0.2\,pc bins were counted and divided by the area of the annuli to obtain the number of filaments per unit area that is shown in Fig.\,6. Even though a skeleton may 
stretch between bins, they were counted only once and do not overlap between bins, because we consider the radial distance of all skeletons as the average of the distance of all pixels of a given skeleton.

\section{Interpretations}

\subsection{Coalescing filaments}
 The Mon\,R2 HFS is a target with a high degree of radial symmetry. Analysis of the orientation of the filaments shows that they are radially focussed on the hub centre towards the Mon\,R2 source (see Fig.\,4a and 4b). 
The average deviation $\overline{\delta\theta}$ of the filament angles from the purely radial direction is close to zero near the hub, suggesting that the skeletons are radially aligned to the hub. The mean deviation from the purely radial direction $\sqrt{\overline{{\Delta\theta}^2}}$ is a slowly increasing function of the radius, suggesting that the skeletons are less radially aligned at larger distances from the hub. The visual spiral appearance \citep{TrevinoMorales2019}, which is prominent only within the hub region, may be represented by the dip at $\sim$1\,pc in Fig.\,4a and the change in the curve direction inside the hub region in Fig.\,4b. The region at $\sim$1.5\,pc, where the angles change, is also the radial distance at which the number of filaments has a maximum (Fig.\,5). This may be an indication of the bending of filaments, possibly due to the rotation of the hub with respect to the outer areas, because the density profile is smooth and shows no such breaks at $\sim$1.5\,pc (Fig.\,3).

In a radially symmetric HFS, if all the filaments were to move inwards towards the hub centre, the number density of filaments should be very large inside the hub. In Fig.\,5, the radial variation of filament M/L, together with the number of filaments is shown. The M/L and the number of filaments can be seen to vary in opposite trends. The M/L of filaments 2--3\,pc away from the hub is typically a few and up to 10 \msun\,pc$^{-1}$. From about 
a radial distance of 1--1.5\,pc, the M/L increases smoothly towards the hub, reaching $\sim$ 100 \msun\,pc$^{-1}$ within the central 0.4\,pc radius 
(the core of the hub) where the curve flattens out.  However, the number of filaments decreases towards the centre of the hub at distances $r \sim$1--1.5\,pc. In Fig.\,5 the number of filaments represents the number of skeletons that intersect concentric circles drawn around the hub as shown in 
Fig.\,A.2. This decrease in filament numbers and increase in M/L towards the hub suggests the coalescence of lower-density filaments into higher-density filaments. If the filaments did not coalesce or if they were in a uniform arrangement of filaments, the number of filaments intersecting the concentric circles would not show the trends of Fig.\,5. A variation of Fig.\,5 where the number of filaments per unit area (N), counted in 0.2\,pc bins, is plotted in Fig.\,6. Viewing from a larger to a smaller radius, here again, the M/L and N increase with decreasing radius from 3\,pc to 1\,pc; however, within 1\,pc of the hub, N flattens out while M/L continues to increase 
towards the centre, suggesting coalescence.

As seen in Figs.\,5 \& 6, the effect of an increase in M/L and a reduction in the number of filaments is prominent within the hub-core of radius 0.4\,pc. At a radius of 1-2\,pc, a large number of low M/L filaments are found to be concentrated which can be seen as a peak in the number of filaments. There is a hint of a second peak in the number of filaments at ~0.6-0.8\,pc (depending on the intervals of concentric circles used to produce the curve), which may be indicative of coalescence taking place in two steps in this target.



The detected filaments appear in a range of lengths and column densities all of which are radially converging to the central hub, where the column 
density reaches its peak values (Fig.\,2a). The white cone-like boundaries were grouped by eye to identify groups of lower-density filaments that appear to culminate at the tip of the cone which marks the beginning of a higher-density filament. Within these groups of filaments, there are junctions with higher column densities, which are marked with red and blue circles. Each of the groups marked by the white boundaries represents cascades of lower-density filaments which appear to funnel into the higher-density filaments, which in turn funnel into even higher-density filaments. We interpret that the coalescence is taking place in at least two steps. The white cones represent the first step of coalescence, and an example of the second step is marked by the black cone in Fig.\,2a. From Fig.\,2b, these steps are reflected in the three density ranges represented by the greenish yellow colour that leads into red-coloured skeletons which in turn funnel into the short white skeletons at the centre.  At each step, these colour changes can be read out as changes in column density that are found to 
increase by a factor of $\sim$5--10 as the skeleton length decreases. In Fig.\,A.3, we display radial profiles of the azimuthally averaged column density within three groups (marked in Fig.\,A.2) of coalescing filaments. These profiles show several jumps, but in all three profiles, the 
slopes change once at 0.8\,pc and again at 0.2\,pc. At larger radii, the changes are difficult to discern, however, groups 1 and 3 display a common bump at $\sim$1.4\,pc. The two peaks in the number of filaments at $\sim$1.5\,pc and at 0.8\,pc in Fig.\,5 likely correspond to the changes apparent in Fig.\,A.3. The net change in surface density along these cascades is from $\sim$5$\times$10$^{20}$cm$^{-2}$ to $\sim$2$\times$10$^{23}$cm$^{-2}$ in the densest portion at the centre just below the Mon\,R2 B-type star. The density jumps become prominent in the direction of the radial 
converging point inside the central hub ($r = $0.8\,pc). In Fig.\,3, the steep and smooth rise in the filament component as a function of radius 
shows that the column density of filaments is gradually increasing towards the centre, at least in the inner 2.5\,pc radius of the cloud. This is in agreement with the description of cascades of low column density filaments funnelling towards higher-density filaments.

Given that the filament networks are observed in projection, not all aligned features are coalescing.  The identified groups are the regions, where the observed column density of the coalesced filaments are roughly equal or larger than the sum of the column densities of the filaments that are merging. The containment of the filaments within the sheet-like cloud enhances the probability of viewing several coalescing junctions, as evidenced by the measurable density jumps that are marked by circles in Fig.\,2a. The incidence of filament merging may be enhanced due to the radial alignment towards the hub.  Hence, based on the radial variations of the number of filaments and M/L, and that of column density junctions and filament angles, the analysed data are best described by coalescence of molecular filaments in Mon\,R2.

\subsection{Hub-filament system mimicking global collapse}

\begin{figure*}[ht]
\centering
\includegraphics[width=\linewidth]{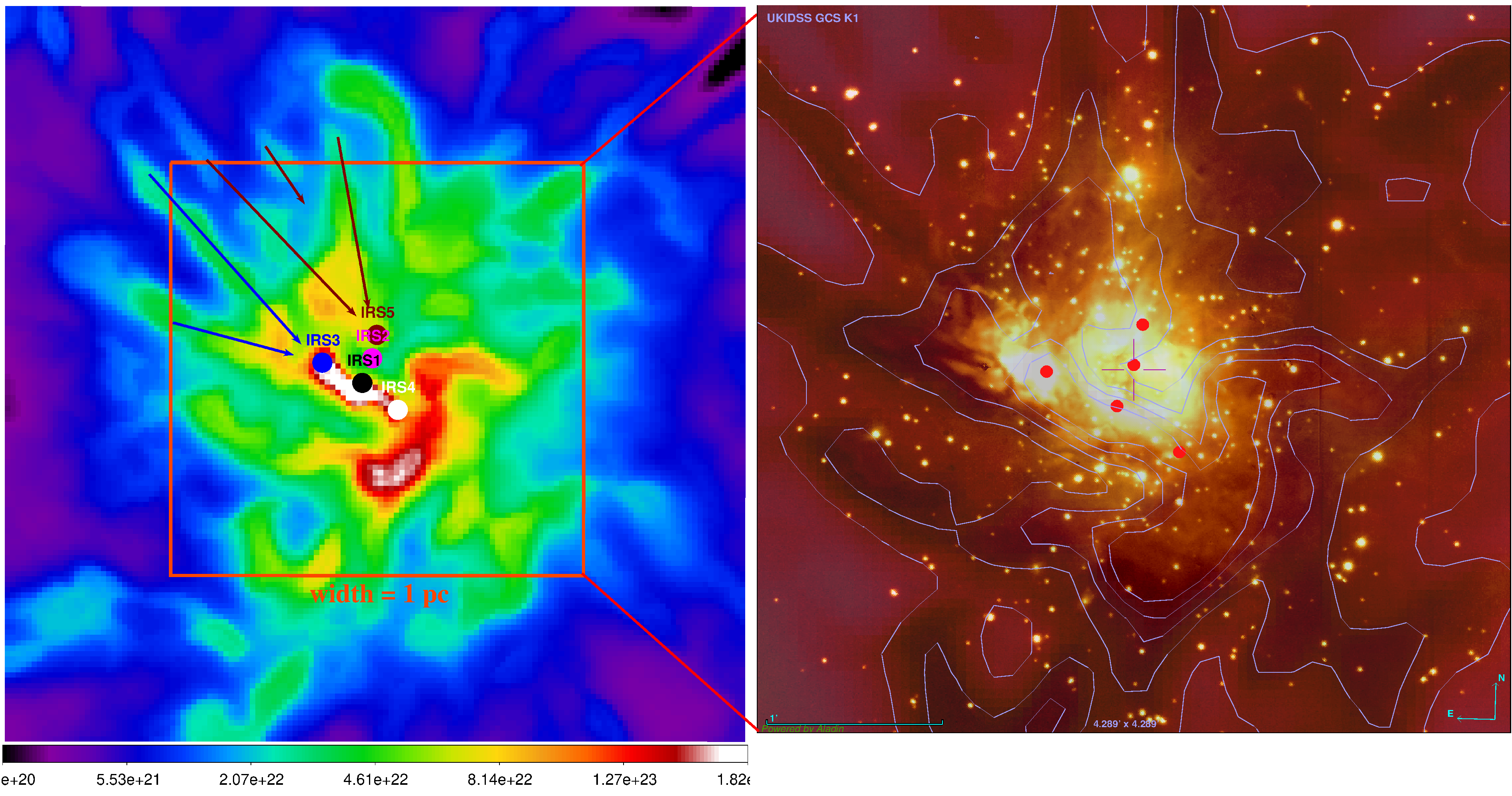}
\caption{Hub is a high-density filament network along which massive stars form. Left: Column density of the filament component of the hub region 
at 11.4\arcsec\, resolution. The most massive sources known in the region, namely IRS1, IRS2, IRS3, IRS4, and IRS5, are marked. The blue and maroon arrows indicate the filaments leading to the junctions in which IRS3 and IRS5 are embedded,  respectively. The red square corresponds to the cluster 
(radius = 0.5\,pc) \citep{carpenter1997}, which is zoomed in on in the right panel. Right: Deep $K$ band (2\mum\,) image overlaid with contours of the 
filament column density image shown in the left panel. Readers should notice the striking correspondence between the infrared dark patches and the column density contours. The IRS sources are marked by red dots.}
\label{fig:massivestars}
\end{figure*}

Fig.\,3 shows the averaged radial column density profile of the observed image, filaments, and diffuse components in a 5\,pc$ \times$ 5\,pc projected region centred on the Mon\,R2 hub. The annular region excluding the hub ($r$ = 0.8\,pc) and extending up to 2\,pc can be well fitted by power-law functions. The column density of the filaments and diffuse cloud components can be fitted by N$_{H_2} \propto r^{-2.17}$ and N$_{H_2} \propto r^{-0.38}$, respectively, where $r$ is the radius from the hub centre, whereas the observed image shows an N$_{H_2}\propto r^{-1.35}$ relation. Sources contribute a small fraction to the total column density (mass fraction of 9--15\%, see Table.\,1), so they are not plotted. The line-of-sight averaging affects the measured column densities and temperatures, and therefore intrinsically one expects lower temperatures and higher densities \citep{Roy2014}. Therefore, the radial profile of the surface density of the filament component may be steeper than the observed N$_{H_2} \propto r^{-2.17}$ relation, but not shallower. Given that the filaments in Mon\,R2 are constrained in a sheet-like volume, almost in the plane of the sky, the observed filament N$_{H_2}$ profile represents the volume density profile $\rho(r)$. Such profiles are indicative of an isothermal sphere in 
equilibrium or at the verge of collapse \citep{Wardthompson94}. Therefore 
it appears that the dense gas (filamentary) component of the hub-filament 
system, implying a $\rho(r)$ profile steeper than $r^{-2}$, is mimicking the conditions for gravitational collapse on parsec scales. 

Such a result has been previously suggested by the velocity gradients \citep{Hacar17b} observed in the Orion ISF, especially towards the BN/KL region. The directional orientation of the velocity gradients towards the gravitational potential of the hub led those authors \citep{Hacar17b,Hacar2018} to suggest that the global collapse scenario may be at work in the Orion ISF. Such behaviour is in some ways similar to that expected from the models of global hierarchical collapse \citep{Enrique2017}. This global 
collapse represents motions of overall dense gas in the hub-filament system towards the hub, similar to the coalescing filaments moving towards the hub in Mon\,R2 (see Sect.\,4.5). This collapse is different from the accretion of matter through longitudinal flows \citep{Peretto14, Williams18} responsible for replenishing the hub, as represented in the F2C model of \citet{Kumar2020}. However, both scenarios suggest a flow of material towards the hub, which is of interest for the formation of massive stars in the hub. This prompts us to estimate the fraction of material in an HFS that is held in the filaments.

\subsection{Filament to hub mass ratio}

 As shown in Sec.\,3.2.2, the filament to hub mass ratio is roughly unity. 
This estimate, however, depends on the tracer used to calculate the mass. 
This is evident from Table\,2 of \citet{TrevinoMorales2019}, where the estimates using CO lines suggest that the hub, which is at a radius of 1\,pc by their definition, holds a significantly lower mass (1600\msun) compared to the filaments (3200\msun) suggesting a filament:hub ratio of 2:1. On the contrary, their dust estimates show that this ratio is inverted with filament (2500\msun):hub (3600\msun) of 1:1.4  \citep{TrevinoMorales2019}. The lower mass estimate from CO lines may be a direct consequence of the line opacity and the removal of gas from the warmer and denser core of the hub. The filament reservoir is important for massive star formation 
because massive stars are known to derive mass through longitudinal flows. If the currently B-type young star (IRS\,1) in the Mon\,R2 core region were to accrete further mass at rates of 10$^{-3}$-10$^{-4}$\msun yr$^{-1}$, as estimated from the observed longitudinal flows \citep{TrevinoMorales2019}, it can grow to become an O star, according to the F2C scenario 
\citep{Kumar2020}. 

In measuring the longitudinal flow rates, \citet{TrevinoMorales2019} conclude that the main (higher-density) filaments transport mass to the central hub at a rate that is four times higher than the secondary (lower-density) filaments to the main filaments. If we consider the coalescence effect in the radial direction, where the main filaments are composed of higher-density structures, that implies that longitudinal flow rates may be simply proportional to the density of the filament. Comparing the annuli in 
Table\,1, the relative mass held in the dense filaments is the highest in 
the hub and slowly declines away from the hub. The diffuse cloud displays 
a reverse trend.  Because the column density correlates with star formation efficiency, the number of sources also (naturally) follows the trend of the dense gas filaments. Sources in these column density maps trace both pre-and proto-stellar cores. The listed mass fraction (\%) of sources in Table\,1 is, therefore, a rough indicator of the core+star formation efficiency within the respective annular regions. It can be seen that the source formation efficiencies in the central hub-core with a radius of 0.4\,pc and in the annular region of 1.6--2.5\,pc radius are 3.4\% and 1.5\%, 
respectively.

\subsection{Structure of hubs and formation of massive stars}

\citet{Myers2009}, in his original definition of hubs, identified them as 
high-column density low-aspect-ratio objects similar to star-forming clumps. Even after several subsequent higher-resolution observations \citep{schneider2012,Peretto14,Williams18,Chen19,TrevinoMorales2019,Kumar2020}, the view of the hub as a massive dense clump has remained. This is because the farther away ($>$ 1\,kpc) objects have mostly been studied by higher spatial resolution observations and the relatively nearby region Mon\,R2 has been studied with beams of 36\arcsec, before this work. From Figs.\,1\,and\,2, and subsequent identification of the hub from Fig.\,3 as the region with a radius of 0.8\,pc, it is found to be a complex network of short, intertwined,  high-column density filaments that coincides with the embedded young stellar cluster. 

This new view of the hub structure has important consequences for our understanding of high-mass star formation. Even though matter can be transported and fed to the stars inside the hub through longitudinal flows, in our new view, whether a star will benefit from it and gain mass depends on 
whether the stellar seed is embedded inside one of the dense filaments within the hub network. Therefore, only some, and not all, of the stellar seeds within the hub can become massive stars. This hypothesis is already supported by the existing data. In Fig.\,7a, we display the filaments component map at 11.4\arcsec resolution, overlaid with the most massive and luminous sources known in Mon\,R2, namely IRS1--5 \citep{hackwell82,Henning92}. The positions for some sources were updated using newer astrometry from the Submillimeter Array (SMA) observation of the central region \citep{Dierickx2015}. The central one-parsec region shown by the red circle in Fig.\,7a is zoomed in on in Fig.\,7b, where a $K$-band image from the UKIDSS \citep{Lawrence2007} Galactic Cluster Survey (GCS) is overlaid with contours of the column density from the image on the left. It is evident from these images that the most massive sources are located on the filament network grid and not in the cavities between them. The luminosities of IRS1, IRS2, IRS3, IRS4, and IRS5 sources are 2000, 100, 3000, 800, and 300 \lsun\, respectively \citep{hackwell82,Henning92,Didelon2015}. Both IRS1 (the well-known B star) and IRS2 (bright $K$-band star) are continuum sources displaying free-free emission. IRS3 is the highest-luminosity object and a previously known hot molecular core \citep{Boonman2003}, driving a powerful massive outflow \citep{Dierickx2015}. IRS5 is also a young object with a rich chemistry  \citep{Dierickx2015}. IRS4 is the lowest-luminosity object in the region. IRS1 is located at the highest column density point in the region. IRS3 and IRS5 are the youngest and most massive sources, located in two separate coalescing junctions identified by the blue and 
maroon coloured arrows in Fig.\,7a. IRS2 is at the centre of the ring-like cavity seen in the $K$-band image, which corresponds to the compact HII 
region of Mon\,R2. 

The locations of the massive sources above are well represented by the F2C scenario \citep[see Fig.\,14 of][]{Kumar2020}. Although the results in this paper move a step further from the simplistic 
two-filament case depicted by the F2C model, the highest-density filament of the region (white filament in Fig.\,7a) is a junction of the filaments marked by the blue arrows. IRS1 and IRS3 represent the older and the younger pair of massive sources from the F2C model, both competing for the same source of longitudinal flow. IRS1 in reality represents the massive star embedded in the white filament (Fig.\,7a), corresponding to the optically visible B star which is revealed through a hole punched out in the densest filament. The ring-shaped compact \ion{H}{II} region in Fig.\,7b is driven by this B star. IRS3 is the most luminous object of the region, located in the exact junction leading to the densest filament. It drives a massive outflow, and therefore it may at present be accreting through the direct longitudinal flows fed through the filaments represented by blue arrows. It is interesting to note that IRS1, IRS3, and IRS4 are equally spaced in the densest filament and therefore the basic stellar seeds
may be the result of filament fragmentation. However, the evolution of these seeds to become a massive star depends upon a competition for the mass reservoir that is present within the densest filament itself and which is arriving through the longitudinal flows. 

The network of intertwined filamentary structures within the hub enhances 
the efficiency of feedback dispersal. The lower column density material between the filaments (with a radial profile of $\Sigma \propto r^{-0.38}$, Fig.\,3) allow the ionising pressure from the young massive stars that formed in the dense filaments of the grid-like hub to escape out of the system 
\citep{Dale11, Kumar2020}. At the same time, the matter accretion continues along the filaments feeding the stellar seeds of the densest filaments 
in the hub that can evolve into massive stars. Together with its network-like structure, the key to 'holding off' massive star formation until the end may come from the magnetic support that threads the dense filaments of the network (see Sec.\,4.5). This allows the build-up of massive cores in very dense filaments and supports them from collapsing until high densities are reached so that when the core collapses, it can begin massive star formation and benefit from additional high accretion rates from longitudinal flows to grow further. In the hub, the network of the filaments has a range in densities, albeit high, which can increase with time, and that can lead to a range in the mass of the massive cores. Depending on the structure of the network, longitudinal flows feeding these massive cores can also differ. These differences may be the factor responsible for producing groups of OB stars with a spectrum of masses in the hub. A good example would be the Trapezium cluster in Orion, with four stars in the range 15--30 \msun that could have formed from an intertwined network of dense filaments similar to that observed in Mon\,R2.

\begin{figure}[ht]
\centering
\includegraphics[width=\linewidth]{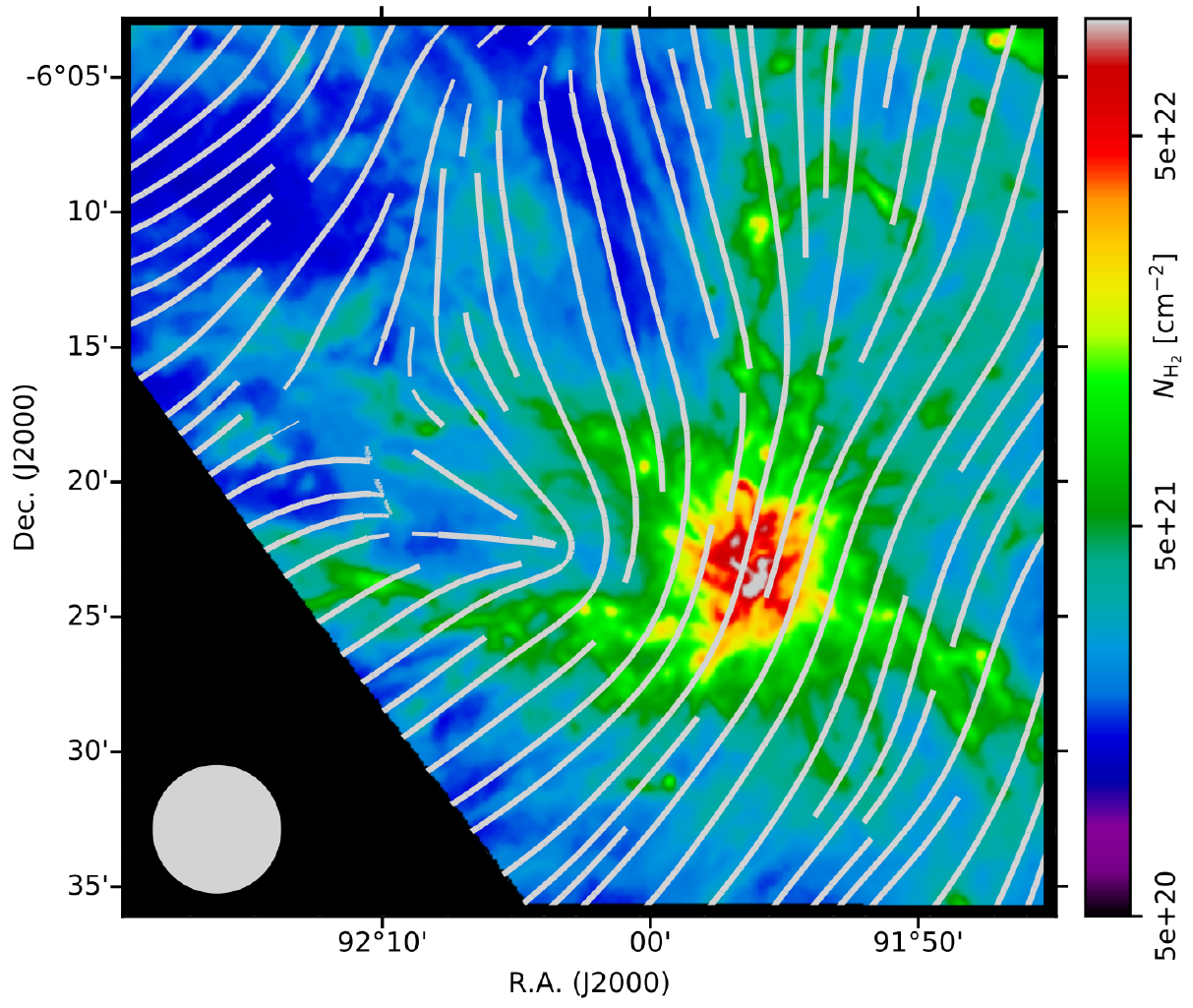}
\caption{\herschel column density map at 18.2\arcsec resolution overlaid by grey streamlines displaying the plane of the sky B fields which were computed using the 353 GHz observations from the \planck mission. The spatial resolution of the \planck data is 4.9\arcmin (as shown by the grey circle in the lower left of the image). The Stokes $Q$ and $U$ components were used to obtain the B-field position angles. The results are displayed as grey streamlines by using \textsl{streamplot} in \textsl{matplotlib}, as thicker lines for the signal-to-noise ratio, S/N $>$ 2.5, and thinner ones for 
2.0 $<$ S/N $<$ 2.5.}
\label{fig:planck}
\end{figure}

\subsection{Magnetic fields: Critical support against global collapse}

The apparent global collapse inferred from the radial dependence of the densities (Sect.\,4.2) roughly extends to a diameter of 4-5\,pc in Mon\,R2, while the velocity gradients found in the central regions of OMC1 \citep{Hacar17} extends to about 2\,pc size. In both cases, the signatures of collapse are directed towards a hub located in a larger molecular cloud complex. These examples may indeed be representative of a more general scenario extending to other star-forming regions with prominent HFS. Evidence presented by \citet{TrevinoMorales2019} suggests that Mon\,R2 in all likelihood is a flattened sheet-like structure. If the OMC\,1 also represents a sheet-like structure viewed edge-on, it is possible that the results 
here and that presented by \citet{Hacar17}, are pointing to a similar phenomenon. In Mon\,R2, the radial orientations of the filaments and the collapse features over scales of 4-5\,pc may be viewed through multiple scenarios, including for example: a) formation of HFS in a sheet-like cloud compressed by a propagating shock front, with filaments converging towards the hubs \citep{Inoue18}, or b) two clouds that are part of a larger virialised cloud complex colliding with each other resulting in filament formation, and the lateral contraction of the compressed layer in an inside-out manner 
leading to radial alignment of the filaments and filament coalescence \citep{Balfour15,Balfour17}. In the F2C model \citep{Kumar2020}, stage I requires filament overlaps to take place to form hubs, allowing all possible scenarios such as a) and b) above to operate. It should be noted that the observed collapse features, both in Mon\,R2 and OMC1, are confined to the inner regions of a much larger sheet-like cloud, and not to the entire cloud, as may be misunderstood by the term 'global collapse'. 

In Mon\,R2, the central region studied here is represented by star formation efficiencies of 5-10\% \citep{Pokhrel16}. This was obtained by assuming a gas-to-star density relation \citep{Gutermuth11}, which is very similar to other nearby star-forming regions \citep{Pokhrel16,Pokhrel20} in Gould's belt. More interestingly, the observed Kennicutt-Schmidt-like relation in this cloud and other clouds has been argued to be linked to a tight linear correlation between the star formation rate per unit area and their gas surface density normalised by the gas free-fall time \citep{Pokhrel21}. The 5-10\% efficiency in Mon\,R2 is higher than the global cloud-scale star formation efficiencies that are typically 2-3\% and free-fall times of 0.5-2\,Myr for nearby molecular clouds \citep{Evans14,Lada13}. However, if the gas reservoir in these collapsing central regions is turned 
into stars with an efficiency similar to that of core-to-star efficiency \citep{Alves2007}, the expected star formation efficiencies are $\sim$30-50\% which would be much higher than the estimated 5-10\%. Given that such high star-formation rates implied by accelerated collapses are unknown, not only in 
Mon\,R2 but for all well-studied nearby star-forming regions, it appears that despite the observed signatures, the HFS is at least partially supported against the global collapse. Such support can be provided by either higher turbulence or magnetic fields. The observed line widths of various 
molecules in both Mon\,R2 \citep{TrevinoMorales2019} and Orion ISF \citep{Hacar2018} do not suggest any unusual non-thermal support against such a 
global collapse. The non-thermal support is minimal (filaments in virial equilibrium \citep{Arzoumanian2013}) in these high-density regions, as reported by the observations of fibres in the Orion ISF \citep{Hacar2018}. Therefore, the critical support needed against the global collapse of the 
Mon\,R2 HFS must come from magnetic fields. Here, we suggest a simplistic 
scenario depicting the role of magnetic fields. Assuming that the B field 
is frozen into the filaments quasi-statically, as the filaments come together and coalesce if the field also merges and reorganises itself, it can provide the necessary support against the global collapse. This scenario is at the outset well represented by the \planck observations (Fig.\,8).

 Throughout much of the region, the B fields are orthogonal in most filaments, and not so much in a few, for example the northern main arm (Fig.\,8). Because the fields are projected onto the sky (2D) and obtained with 
a large beam of 4.9\arcmin, we may expect relatively more inclined projections for some places in Mon\,R2, with its inclination of 30\deg \ to the line of sight. Results prior to the \planck mission using starlight polarisation \citep{Palmeirim13}, from \planck \citep{planck2016-a,planck2016-b,planck2016-c} and those combining the starlight and dust polarisation \citep{cox16}, show that B fields are parallel to low-column density filaments and perpendicular to high-column density filaments. More recent observations (after 2016) at higher angular resolutions, using JCMT and SOFIA facilities \citep{Pattle19,Fabio19,Pillai20,Arzoumanian2021}, show 
that this is not exactly the case, especially for high-column density filaments. The B field inside the filaments (when resolved) shows some variations and it is not always perpendicular. \citet{Pillai20} also found similar results to the post \planck studies and suggest the presence of two transitions, one from parallel to perpendicular at A$_v \sim$7-8\,mag (similar to \planck results \citet{soler16}) and a transition from perpendicular to parallel for A$_v >$21\,mag, where the B field was dragged by the flow and gravity along the filaments, compatible with the simulations of \citet{Gomez18}.  
 
 In Mon\,R2, the filaments identified by our analysis are all well above A$_v \sim$7\,mag and they coalesce to form higher column density filaments towards the hub. Therefore, to the first order, all dense filaments have 
magnetic fields perpendicular to their axes in Fig.\,8, and less orthogonal orientations between field lines and filaments can be viewed as effects of projection. If the B fields are frozen to the filaments quasi-statically, then, as the filaments coalesce and move towards the radial centre of the hub,  
they come closer, resulting in the interaction of B fields frozen to individual filaments. This can lead to field distortions and reorganisations naturally. The dip in the field lines between the two eastern arms in Fig.\,8 may be hinting at such interactions. Although the B field may be dragged by flows within some filaments, it is not a ubiquitous feature and 
depends upon local conditions and the strength of longitudinal flows. On the contrary, the frozen-in-fields with a largely orthogonal orientation to 
the dense filaments are ubiquitously observed \citep{Palmeirim13,cox16, soler16}. It is unknown if the B fields are smoothly connected between the 
filaments throughout the cloud, such as the results by many numerical simulations \citep[e.g.][]{Gomez18, Inoue18}. Given that the nature of the magnetic field in the low-density regions between filaments is unknown so far, field interaction of the frozen-in fields between filaments must be investigated. Mon\,R2 is an excellent target to study if the changes in field direction take place because of the flow-induced entrainments, field 
interactions, or a combination of both. 

The magnetic field structures within the hub regions are relatively uncertain, especially in the light of the results that the hub is a network of 
short dense filaments, as opposed to the idea of a uniform clump. B-field 
mapping of the hub with higher spatial resolutions can shed light on its configuration. However, we suggest that the overall field direction would 
result in hourglass shapes, perpendicular to a flattened hub (a network of high-density filaments in a sheet-like space). This inference is guided 
by comparisons of the observations of known HFS. In the nearby Ophiuchus region \citep{Pattle19, Fabio19}, the polarisation data provide a spatial resolution of $\sim$0.01\,pc. If we view Oph\,A as an edge-on hub 
in the shape of an arc, then the magnetic fields are oriented perpendicular to the arc \citep[Fig.\,2 of][]{Fabio19}. In NGC6334 \citep{Arzoumanian2021}, the average B field runs perpendicular to the main clump-hub at 1\,pc scale (similar in size to the hub in Mon\,R2), and a similar situation is in the main hub of Serpens \citep{Pillai20}. Such field configurations of the hubs are crucial to provide strong support against the collapse 
of the hub itself, instead of allowing it to accrete large amounts of material from the surrounding filaments (a secondary reservoir in the F2C model), therefore aiding the formation of massive stars within the hub \citep{Kumar2020}.

\section{Discussion}

Returning to the original questions posed in Sec.\,1, namely, what the 
difference is between bundles of fibres forming higher M/L filaments and a hub, which is a junction of  filaments. It appears that fibre and filament coalescence is the key mechanism that operates in both circumstances. Essentially, fibres and filaments of gas and dust in a dynamical molecular cloud coalesce in different ways, depending on the external influencing factors, such as shock waves and gravitational potentials. In this coalescence process, it appears that the angle of incidence between filaments forming the junction and the gravitational potential difference across the junction control the final product. For example, acute angles are more likely to produce high linear-density (high M/L) filaments, while obtuse angles may result in hubs. If one were to suppose that there are competing centres of 
gravitational potential formed by two hubs located a few parsecs away, the potential gradients would impact how the coalescing structures evolve. Similarly, if a shock wave were to sweep the filaments into a sheet, or be compressed in a layer during cloud-cloud \citep{InoueFukui13,Fukui21} or bubble-bubble \citep{Inutsuka15} collision, it may enhance the formation of 
bundles and high M/L filaments. Elaborate studies are required to study and understand these dynamic conditions. 

For example, \citet{Hacar17} note that the super-critical fibres in NGC1333 survive for roughly five free-fall times as independent entities, which they attributed to the support provided by non-thermal motions. In contrast, the dense network of fibres observed in the Orion ISF \citep{Hacar2018} is essentially located within the boundaries of the hub (radius of 0.8\,pc), similar to that in Mon\,R2. By their very nature, hubs are junctions that formed as a consequence of dynamical evolution within the cloud; therefore, they are younger structures compared to the rest of the cloud. So, it is unclear how long such bundles of velocity coherent fibres (near or within the hub) can survive before the internal velocity differences diffuse and the bundle itself becomes velocity coherent leading to a coalesced high-linear-density filament collapsing onto its axis and fragmenting to form stars \citep{Inutsuka1997}. 

The observational features that mimic global collapses, such as the radial density profiles in Mon\,R2 or the steep velocity gradients close to the hub in Orion ISF \citep{Hacar17}, imply that the gravitational potential in these HFS is significant and influences the surrounding material up 
to at least 2\,pc. In Mon\,R2, this hub potential may be responsible for the bulk contraction of the central cluster as evidenced by kinematic data of stars (albeit with large errors) \citep[see Table\,2 of][]{Kuhn2019}. Also, the radial focussing of the filaments found here is driven by the gravitational potential of the hub. If the HFS are in a 2D geometrical configuration of a layer, both the dense and diffuse components of the clouds will be compressed into the layer. The hub's significant gravitational potential can then even lead to the formation of further filaments close to the hub, by streaming the diffuse cloud component along the radially directed gravitational field. 

When fibres and filaments coalesce to form a more spherical and circular structure, whether it remains that way or evolves into a different structure appears to depend on internal density gradients and external forces. In Mon\,R2, as marked by the red and blue circles in Fig.\,2a, there are several junctions with density enhancements of low aspect ratios, satisfying the definition of \citet{Myers2009} to be referred to as a hub. However, these junctions in Mon\,R2, especially those that are exterior to the hub, have 
strong internal density gradients, that are directed towards the hub; therefore, they are more likely transitory in the sense that they are the nodes feeding higher-linear-density filaments, and moving towards the main 
hub along with other coalescing structures. This detailed view of the junctions in Mon\,R2 and also of those observed in NGC1333 and Orion ISF suggests that they may have a range in sizes, depending on the number and density of the filaments forming the junction. The top-level structures resulting from a hierarchical coalescence in a dynamically evolving cloud correspond to the main hubs as originally defined in nearby clouds by \citet{Myers2009} and represented by the HFS clumps studied by \citet{Kumar2020}.

In its simplest form, the volume density of cold dense matter controls the process of star formation, consequently influencing the key observational
properties, as often demonstrated by the studies of the probability density functions \citep{jounipdf}, such as rho-PDF (volume) and N-PDF (surface). Filament coalescence may be a key factor in assembling the density spectrum over several dynamical scales in the lifetime of molecular clouds. During this time, the density functions can also be influenced by the gravitational potential of clumps and hubs from within, and by 
external shock waves that could restructure the cloud material, for example into sheets, enhancing the efficiency of coalescence, as in the case of Mon\,R2.

\begin{acknowledgements}
    We thank the referee for critical and thorough scrutiny that greatly enhanced the flow of the manuscript and improved the clarity of the presentation. MSNK thanks Prof.\,Anthony Whitworth for stimulating discussions 
on sheets and cloud-cloud collisions. MSNK acknowledges the support from  
FCT  -  Funda\c{c}\~{a}o para a  Ci\^{e}ncia  e  a  Tecnologia  through  Investigador contracts and exploratory project   (IF/00956/2015/CP1273/CT0002). PP   acknowledge support from   FCT/MCTES   through   Portuguese national funds   (PIDDAC) by grant    UID/FIS/04434/2019. PP  receives support from fellowship SFRH/BPD/110176/2015 funded by FCT(Portugal) and POPH/FSE(EC).  M. M. is supported by JSPS KAKENHI grant No. 20K03276. The \herschel data were obtained from the Herschel Science Archive (HSA) and were obtained through the guaranteed time key programmer HOBYS (PI: F. Motte). We thank Pierre Didelon for providing the calibration offsets that were necessary to obtain flux measurements from PACS data.
\end{acknowledgements}

%
%
\bibliographystyle{aa}
\bibliography{40363corr}

\begin{thebibliography}{71}
\expandafter\ifx\csname natexlab\endcsname\relax\def\natexlab#1{#1}\fi

\bibitem[{{Alves} {et~al.}(2007){Alves}, {Lombardi}, \& {Lada}}]{Alves2007}
{Alves}, J., {Lombardi}, M., \& {Lada}, C.~J. 2007, \aap, 462, L17

\bibitem[{{Anderson} {et~al.}(2012){Anderson}, {Zavagno}, {Deharveng},
  {Abergel}, {Motte}, {Andr{\'e}}, {Bernard}, {Bontemps}, {Hennemann}, {Hill},
  {Rod{\'o}n}, {Roussel}, \& {Russeil}}]{Anderson2012}
{Anderson}, L.~D., {Zavagno}, A., {Deharveng}, L., {et~al.} 2012, \aap, 542,
  A10

\bibitem[{{Andr{\'e}} {et~al.}(2019){Andr{\'e}}, {Arzoumanian}, {K{\"o}nyves},
  {Shimajiri}, \& {Palmeirim}}]{Andre19}
{Andr{\'e}}, P., {Arzoumanian}, D., {K{\"o}nyves}, V., {Shimajiri}, Y., \&
  {Palmeirim}, P. 2019, \aap, 629, L4

\bibitem[{{Andr{\'e}} {et~al.}(2014){Andr{\'e}}, {Di Francesco},
  {Ward-Thompson}, {Inutsuka}, {Pudritz}, \& {Pineda}}]{Andrepp6}
{Andr{\'e}}, P., {Di Francesco}, J., {Ward-Thompson}, D., {et~al.} 2014, in
  Protostars and Planets VI, ed. H.~{Beuther}, R.~S. {Klessen}, C.~P.
  {Dullemond}, \& T.~{Henning}, 27

\bibitem[{{Andr{\'e}} {et~al.}(2016){Andr{\'e}}, {Rev{\'e}ret}, {K{\"o}nyves},
  {Arzoumanian}, {Tig{\'e}}, {Gallais}, {Roussel}, {Le Pennec}, {Rodriguez},
  {Doumayrou}, {Dubreuil}, {Lortholary}, {Martignac}, {Talvard}, {Delisle},
  {Visticot}, {Dumaye}, {De Breuck}, {Shimajiri}, {Motte}, {Bontemps},
  {Hennemann}, {Zavagno}, {Russeil}, {Schneider}, {Palmeirim}, {Peretto},
  {Hill}, {Minier}, {Roy}, \& {Rygl}}]{Andre16}
{Andr{\'e}}, P., {Rev{\'e}ret}, V., {K{\"o}nyves}, V., {et~al.} 2016, \aap,
  592, A54

\bibitem[{{Arzoumanian} {et~al.}(2011){Arzoumanian}, {Andr{\'e}}, {Didelon},
  {K{\"o}nyves}, {Schneider}, {Men'shchikov}, {Sousbie}, {Zavagno}, {Bontemps},
  {di Francesco}, {Griffin}, {Hennemann}, {Hill}, {Kirk}, {Martin}, {Minier},
  {Molinari}, {Motte}, {Peretto}, {Pezzuto}, {Spinoglio}, {Ward-Thompson},
  {White}, \& {Wilson}}]{Arzoumanian11}
{Arzoumanian}, D., {Andr{\'e}}, P., {Didelon}, P., {et~al.} 2011, \aap, 529, L6

\bibitem[{{Arzoumanian} {et~al.}(2019){Arzoumanian}, {Andr{\'e}},
  {K{\"o}nyves}, {Palmeirim}, {Roy}, {Schneider}, {Benedettini}, {Didelon}, {Di
  Francesco}, {Kirk}, \& {Ladjelate}}]{Arzoumanian2019}
{Arzoumanian}, D., {Andr{\'e}}, P., {K{\"o}nyves}, V., {et~al.} 2019, \aap,
  621, A42

\bibitem[{{Arzoumanian} {et~al.}(2013){Arzoumanian}, {Andr{\'e}}, {Peretto}, \&
  {K{\"o}nyves}}]{Arzoumanian2013}
{Arzoumanian}, D., {Andr{\'e}}, P., {Peretto}, N., \& {K{\"o}nyves}, V. 2013,
  \aap, 553, A119

\bibitem[{{Arzoumanian} {et~al.}(2021){Arzoumanian}, {Furuya}, {Hasegawa},
  {Tahani}, {Sadavoy}, {Hull}, {Johnstone}, {Koch}, {Inutsuka}, {Doi}, {Hoang},
  {Onaka}, {Iwasaki}, {Shimajiri}, {Inoue}, {Peretto}, {Andr{\'e}}, {Bastien},
  {Berry}, {Chen}, {Di Francesco}, {Eswaraiah}, {Fanciullo}, {Fissel}, {Hwang},
  {Kang}, {Kim}, {Kim}, {Kirchschlager}, {Kwon}, {Lee}, {Liu}, {Lyo}, {Pattle},
  {Soam}, {Tang}, {Whitworth}, {Ching}, {Coud{\'e}}, {Wang}, {Ward-Thompson},
  {Lai}, {Qiu}, {Bourke}, {Byun}, {Chen}, {Chen}, {Chen}, {Cho}, {Choi},
  {Choi}, {Chrysostomou}, {Chung}, {Dai}, {Diep}, {Duan}, {Duan}, {Eden},
  {Fiege}, {Franzmann}, {Friberg}, {Fuller}, {Gledhill}, {Graves}, {Greaves},
  {Griffin}, {Gu}, {Han}, {Hatchell}, {Hayashi}, {Houde}, {Jeong}, {Kang},
  {Kang}, {Kataoka}, {Kawabata}, {Kemper}, {Kim}, {Kim}, {Kim}, {Kim}, {Kirk},
  {Kobayashi}, {K{\"o}nyves}, {Kusune}, {Kwon}, {Lacaille}, {Law}, {Lee},
  {Lee}, {Lee}, {Lee}, {Lee}, {Li}, {Li}, {Li}, {Liu}, {Liu}, {Liu}, {Lu},
  {Mairs}, {Matsumura}, {Matthews}, {Moriarty-Schieven}, {Nagata}, {Nakamura},
  {Nakanishi}, {Ngoc}, {Ohashi}, {Park}, {Parsons}, {Pyo}, {Qian}, {Rao},
  {Rawlings}, {Rawlings}, {Retter}, {Richer}, {Rigby}, {Saito}, {Savini},
  {Scaife}, {Seta}, {Shinnaga}, {Tamura}, {Tang}, {Tomisaka}, {Tram},
  {Tsukamoto}, {Viti}, {Wang}, {Xie}, {Yen}, {Yoo}, {Yuan}, {Yun}, {Zenko},
  {Zhang}, {Zhang}, {Zhang}, {Zhou}, {Zhu}, {de Looze}, {Dowell}, {Eyres},
  {Falle}, {Friesen}, {Robitaille}, \& {van Loo}}]{Arzoumanian2021}
{Arzoumanian}, D., {Furuya}, R.~S., {Hasegawa}, T., {et~al.} 2021, \aap, 647,
  A78

\bibitem[{{Balfour} {et~al.}(2017){Balfour}, {Whitworth}, \&
  {Hubber}}]{Balfour17}
{Balfour}, S.~K., {Whitworth}, A.~P., \& {Hubber}, D.~A. 2017, \mnras, 465,
  3483

\bibitem[{{Balfour} {et~al.}(2015){Balfour}, {Whitworth}, {Hubber}, \&
  {Jaffa}}]{Balfour15}
{Balfour}, S.~K., {Whitworth}, A.~P., {Hubber}, D.~A., \& {Jaffa}, S.~E. 2015,
  \mnras, 453, 2471

\bibitem[{{Barranco} \& {Goodman}(1998)}]{Barranco1998}
{Barranco}, J.~A. \& {Goodman}, A.~A. 1998, \apj, 504, 207

\bibitem[{{Bonne} {et~al.}(2020){Bonne}, {Schneider}, {Bontemps}, {Clarke},
  {Gusdorf}, {Lehmann}, {Steinke}, {Csengeri}, {Kabanovic}, {Simon},
  {Buchbender}, \& {G{\"u}sten}}]{Bonne20}
{Bonne}, L., {Schneider}, N., {Bontemps}, S., {et~al.} 2020, \aap, 641, A17

\bibitem[{{Boonman} \& {van Dishoeck}(2003)}]{Boonman2003}
{Boonman}, A.~M.~S. \& {van Dishoeck}, E.~F. 2003, \aap, 403, 1003

\bibitem[{{Carpenter} {et~al.}(1997){Carpenter}, {Meyer}, {Dougados}, {Strom},
  \& {Hillenbrand}}]{carpenter1997}
{Carpenter}, J.~M., {Meyer}, M.~R., {Dougados}, C., {Strom}, S.~E., \&
  {Hillenbrand}, L.~A. 1997, \aj, 114, 198

\bibitem[{{Chen} {et~al.}(2019){Chen}, {Zhang}, {Wright}, {Busquet}, {Lin},
  {Liu}, {Olguin}, {Sanhueza}, {Nakamura}, {Palau}, {Ohashi}, {Tatematsu}, \&
  {Liao}}]{Chen19}
{Chen}, H.-R.~V., {Zhang}, Q., {Wright}, M.~C.~H., {et~al.} 2019, \apj, 875, 24

\bibitem[{{Clarke} {et~al.}(2018){Clarke}, {Whitworth}, {Spowage},
  {Duarte-Cabral}, {Suri}, {Jaffa}, {Walch}, \& {Clark}}]{clarke18}
{Clarke}, S.~D., {Whitworth}, A.~P., {Spowage}, R.~L., {et~al.} 2018, \mnras,
  479, 1722

\bibitem[{{Cox} {et~al.}(2016){Cox}, {Arzoumanian}, {Andr{\'e}}, {Rygl},
  {Prusti}, {Men'shchikov}, {Royer}, {K{\'o}sp{\'a}l}, {Palmeirim}, {Ribas},
  {K{\"o}nyves}, {Bernard}, {Schneider}, {Bontemps}, {Merin}, {Vavrek}, {Alves
  de Oliveira}, {Didelon}, {Pilbratt}, \& {Waelkens}}]{cox16}
{Cox}, N.~L.~J., {Arzoumanian}, D., {Andr{\'e}}, P., {et~al.} 2016, \aap, 590,
  A110

\bibitem[{{Dale} \& {Bonnell}(2011)}]{Dale11}
{Dale}, J.~E. \& {Bonnell}, I. 2011, \mnras, 414, 321

\bibitem[{{Didelon} {et~al.}(2015){Didelon}, {Motte}, {Tremblin}, {Hill},
  {Hony}, {Hennemann}, {Hennebelle}, {Anderson}, {Galliano}, {Schneider},
  {Rayner}, {Rygl}, {Louvet}, {Zavagno}, {K{\"o}nyves}, {Sauvage}, {Andr{\'e}},
  {Bontemps}, {Peretto}, {Griffin}, {Gonz{\'a}lez}, {Lebouteiller},
  {Arzoumanian}, {Bernard}, {Benedettini}, {Di Francesco}, {Men'shchikov},
  {Minier}, {Nguy{\^e}n Luong}, {Palmeirim}, {Pezzuto}, {Rivera-Ingraham},
  {Russeil}, {Ward-Thompson}, \& {White}}]{Didelon2015}
{Didelon}, P., {Motte}, F., {Tremblin}, P., {et~al.} 2015, \aap, 584, A4

\bibitem[{{Dierickx} {et~al.}(2015){Dierickx}, {Jim{\'e}nez-Serra}, {Rivilla},
  \& {Zhang}}]{Dierickx2015}
{Dierickx}, M., {Jim{\'e}nez-Serra}, I., {Rivilla}, V.~M., \& {Zhang}, Q. 2015,
  \apj, 803, 89

\bibitem[{{Evans} {et~al.}(2014){Evans}, {Heiderman}, \&
  {Vutisalchavakul}}]{Evans14}
{Evans}, Neal~J., I., {Heiderman}, A., \& {Vutisalchavakul}, N. 2014, \apj,
  782, 114

\bibitem[{{Fukui} {et~al.}(2021){Fukui}, {Habe}, {Inoue}, {Enokiya}, \&
  {Tachihara}}]{Fukui21}
{Fukui}, Y., {Habe}, A., {Inoue}, T., {Enokiya}, R., \& {Tachihara}, K. 2021,
  \pasj, 73, S1

\bibitem[{{G{\'o}mez} {et~al.}(2018){G{\'o}mez}, {V{\'a}zquez-Semadeni}, \&
  {Zamora-Avil{\'e}s}}]{Gomez18}
{G{\'o}mez}, G.~C., {V{\'a}zquez-Semadeni}, E., \& {Zamora-Avil{\'e}s}, M.
  2018, \mnras, 480, 2939

\bibitem[{{Goodman} {et~al.}(2014){Goodman}, {Alves}, {Beaumont}, {Benjamin},
  {Borkin}, {Burkert}, {Dame}, {Jackson}, {Kauffmann}, {Robitaille}, \&
  {Smith}}]{Goodman14}
{Goodman}, A.~A., {Alves}, J., {Beaumont}, C.~N., {et~al.} 2014, \apj, 797, 53

\bibitem[{{Goodman} {et~al.}(1998){Goodman}, {Barranco}, {Wilner}, \&
  {Heyer}}]{Goodman98}
{Goodman}, A.~A., {Barranco}, J.~A., {Wilner}, D.~J., \& {Heyer}, M.~H. 1998,
  \apj, 504, 223

\bibitem[{{Gutermuth} {et~al.}(2011){Gutermuth}, {Pipher}, {Megeath}, {Myers},
  {Allen}, \& {Allen}}]{Gutermuth11}
{Gutermuth}, R.~A., {Pipher}, J.~L., {Megeath}, S.~T., {et~al.} 2011, \apj,
  739, 84

\bibitem[{{Hacar} {et~al.}(2017{\natexlab{a}}){Hacar}, {Alves}, {Tafalla}, \&
  {Goicoechea}}]{Hacar17b}
{Hacar}, A., {Alves}, J., {Tafalla}, M., \& {Goicoechea}, J.~R.
  2017{\natexlab{a}}, \aap, 602, L2

\bibitem[{{Hacar} {et~al.}(2016){Hacar}, {Kainulainen}, {Tafalla}, {Beuther},
  \& {Alves}}]{Hacar16}
{Hacar}, A., {Kainulainen}, J., {Tafalla}, M., {Beuther}, H., \& {Alves}, J.
  2016, \aap, 587, A97

\bibitem[{{Hacar} {et~al.}(2017{\natexlab{b}}){Hacar}, {Tafalla}, \&
  {Alves}}]{Hacar17}
{Hacar}, A., {Tafalla}, M., \& {Alves}, J. 2017{\natexlab{b}}, \aap, 606, A123

\bibitem[{{Hacar} {et~al.}(2018){Hacar}, {Tafalla}, {Forbrich}, {Alves},
  {Meingast}, {Grossschedl}, \& {Teixeira}}]{Hacar2018}
{Hacar}, A., {Tafalla}, M., {Forbrich}, J., {et~al.} 2018, \aap, 610, A77

\bibitem[{{Hackwell} {et~al.}(1982){Hackwell}, {Grasdalen}, \&
  {Gehrz}}]{hackwell82}
{Hackwell}, J.~A., {Grasdalen}, G.~L., \& {Gehrz}, R.~D. 1982, \apj, 252, 250

\bibitem[{{Heiles}(1979)}]{Heiles79}
{Heiles}, C. 1979, \apj, 229, 533

\bibitem[{{Henning} {et~al.}(1992){Henning}, {Chini}, \& {Pfau}}]{Henning92}
{Henning}, T., {Chini}, R., \& {Pfau}, W. 1992, \aap, 263, 285

\bibitem[{{Inoue} \& {Fukui}(2013)}]{InoueFukui13}
{Inoue}, T. \& {Fukui}, Y. 2013, \apjl, 774, L31

\bibitem[{{Inoue} {et~al.}(2018){Inoue}, {Hennebelle}, {Fukui}, {Matsumoto},
  {Iwasaki}, \& {Inutsuka}}]{Inoue18}
{Inoue}, T., {Hennebelle}, P., {Fukui}, Y., {et~al.} 2018, \pasj, 70, S53

\bibitem[{{Inutsuka} {et~al.}(2015){Inutsuka}, {Inoue}, {Iwasaki}, \&
  {Hosokawa}}]{Inutsuka15}
{Inutsuka}, S.-i., {Inoue}, T., {Iwasaki}, K., \& {Hosokawa}, T. 2015, \aap,
  580, A49

\bibitem[{{Inutsuka} \& {Miyama}(1997)}]{Inutsuka1997}
{Inutsuka}, S.-i. \& {Miyama}, S.~M. 1997, \apj, 480, 681

\bibitem[{{Kainulainen} {et~al.}(2014){Kainulainen}, {Federrath}, \&
  {Henning}}]{jounipdf}
{Kainulainen}, J., {Federrath}, C., \& {Henning}, T. 2014, Science, 344, 183

\bibitem[{{Kuhn} {et~al.}(2019){Kuhn}, {Hillenbrand}, {Sills}, {Feigelson}, \&
  {Getman}}]{Kuhn2019}
{Kuhn}, M.~A., {Hillenbrand}, L.~A., {Sills}, A., {Feigelson}, E.~D., \&
  {Getman}, K.~V. 2019, \apj, 870, 32

\bibitem[{{Kumar} {et~al.}(2020){Kumar}, {Palmeirim}, {Arzoumanian}, \&
  {Inutsuka}}]{Kumar2020}
{Kumar}, M.~S.~N., {Palmeirim}, P., {Arzoumanian}, D., \& {Inutsuka}, S.-i.
  2020, \aap, 642, A87

\bibitem[{{Lada} {et~al.}(2013){Lada}, {Lombardi}, {Roman-Zuniga}, {Forbrich},
  \& {Alves}}]{Lada13}
{Lada}, C.~J., {Lombardi}, M., {Roman-Zuniga}, C., {Forbrich}, J., \& {Alves},
  J.~F. 2013, \apj, 778, 133

\bibitem[{{Lawrence} {et~al.}(2007){Lawrence}, {Warren}, {Almaini}, {Edge},
  {Hambly}, {Jameson}, {Lucas}, {Casali}, {Adamson}, {Dye}, {Emerson},
  {Foucaud}, {Hewett}, {Hirst}, {Hodgkin}, {Irwin}, {Lodieu}, {McMahon},
  {Simpson}, {Smail}, {Mortlock}, \& {Folger}}]{Lawrence2007}
{Lawrence}, A., {Warren}, S.~J., {Almaini}, O., {et~al.} 2007, \mnras, 379,
  1599

\bibitem[{{Men'shchikov}(2021{\natexlab{a}})}]{sasha21}
{Men'shchikov}, A. 2021{\natexlab{a}}, \aap, 654, A78

\bibitem[{{Men'shchikov}(2021{\natexlab{b}})}]{sasha20}
{Men'shchikov}, A. 2021{\natexlab{b}}, \aap, 649, A89

\bibitem[{{Molinari} {et~al.}(2010){Molinari}, {Swinyard}, {Bally}, {Barlow},
  {Bernard}, {Martin}, {Moore}, {Noriega-Crespo}, {Plume}, {Testi}, {Zavagno},
  {Abergel}, {Ali}, {Anderson}, {Andr{\'e}}, {Baluteau}, {Battersby},
  {Beltr{\'a}n}, {Benedettini}, {Billot}, {Blommaert}, {Bontemps}, {Boulanger},
  {Brand}, {Brunt}, {Burton}, {Calzoletti}, {Carey}, {Caselli}, {Cesaroni},
  {Cernicharo}, {Chakrabarti}, {Chrysostomou}, {Cohen}, {Compiegne}, {de
  Bernardis}, {de Gasperis}, {di Giorgio}, {Elia}, {Faustini}, {Flagey},
  {Fukui}, {Fuller}, {Ganga}, {Garcia-Lario}, {Glenn}, {Goldsmith}, {Griffin},
  {Hoare}, {Huang}, {Ikhenaode}, {Joblin}, {Joncas}, {Juvela}, {Kirk},
  {Lagache}, {Li}, {Lim}, {Lord}, {Marengo}, {Marshall}, {Masi}, {Massi},
  {Matsuura}, {Minier}, {Miville-Desch{\^e}nes}, {Montier}, {Morgan}, {Motte},
  {Mottram}, {M{\"u}ller}, {Natoli}, {Neves}, {Olmi}, {Paladini}, {Paradis},
  {Parsons}, {Peretto}, {Pestalozzi}, {Pezzuto}, {Piacentini}, {Piazzo},
  {Polychroni}, {Pomar{\`e}s}, {Popescu}, {Reach}, {Ristorcelli}, {Robitaille},
  {Robitaille}, {Rod{\'o}n}, {Roy}, {Royer}, {Russeil}, {Saraceno}, {Sauvage},
  {Schilke}, {Schisano}, {Schneider}, {Schuller}, {Schulz}, {Sibthorpe},
  {Smith}, {Smith}, {Spinoglio}, {Stamatellos}, {Strafella}, {Stringfellow},
  {Sturm}, {Taylor}, {Thompson}, {Traficante}, {Tuffs}, {Umana}, {Valenziano},
  {Vavrek}, {Veneziani}, {Viti}, {Waelkens}, {Ward-Thompson}, {White},
  {Wilcock}, {Wyrowski}, {Yorke}, \& {Zhang}}]{Molinari2010}
{Molinari}, S., {Swinyard}, B., {Bally}, J., {et~al.} 2010, \aap, 518, L100

\bibitem[{{Motte} {et~al.}(2010){Motte}, {Zavagno}, {Bontemps}, {Schneider},
  {Hennemann}, {di Francesco}, {Andr{\'e}}, {Saraceno}, {Griffin}, {Marston},
  {Ward-Thompson}, {White}, {Minier}, {Men'shchikov}, {Hill}, {Abergel},
  {Anderson}, {Aussel}, {Balog}, {Baluteau}, {Bernard}, {Cox}, {Csengeri},
  {Deharveng}, {Didelon}, {di Giorgio}, {Hargrave}, {Huang}, {Kirk}, {Leeks},
  {Li}, {Martin}, {Molinari}, {Nguyen-Luong}, {Olofsson}, {Persi}, {Peretto},
  {Pezzuto}, {Roussel}, {Russeil}, {Sadavoy}, {Sauvage}, {Sibthorpe},
  {Spinoglio}, {Testi}, {Teyssier}, {Vavrek}, {Wilson}, \&
  {Woodcraft}}]{motte2010}
{Motte}, F., {Zavagno}, A., {Bontemps}, S., {et~al.} 2010, \aap, 518, L77

\bibitem[{{Myers}(2009)}]{Myers2009}
{Myers}, P.~C. 2009, \apj, 700, 1609

\bibitem[{{Palmeirim} {et~al.}(2013){Palmeirim}, {Andr{\'e}}, {Kirk},
  {Ward-Thompson}, {Arzoumanian}, {K{\"o}nyves}, {Didelon}, {Schneider},
  {Benedettini}, {Bontemps}, {Di Francesco}, {Elia}, {Griffin}, {Hennemann},
  {Hill}, {Martin}, {Men'shchikov}, {Molinari}, {Motte}, {Nguyen Luong},
  {Nutter}, {Peretto}, {Pezzuto}, {Roy}, {Rygl}, {Spinoglio}, \&
  {White}}]{Palmeirim13}
{Palmeirim}, P., {Andr{\'e}}, P., {Kirk}, J., {et~al.} 2013, \aap, 550, A38

\bibitem[{{Pattle} {et~al.}(2019){Pattle}, {Lai}, {Hasegawa}, {Wang}, {Furuya},
  {Ward-Thompson}, {Bastien}, {Coud{\'e}}, {Eswaraiah}, {Fanciullo}, {di
  Francesco}, {Hoang}, {Kim}, {Kwon}, {Lee}, {Liu}, {Liu}, {Matsumura},
  {Onaka}, {Sadavoy}, \& {Soam}}]{Pattle19}
{Pattle}, K., {Lai}, S.-P., {Hasegawa}, T., {et~al.} 2019, \apj, 880, 27

\bibitem[{{Peretto} {et~al.}(2014){Peretto}, {Fuller}, {Andr{\'e}},
  {Arzoumanian}, {Rivilla}, {Bardeau}, {Duarte Puertas}, {Guzman Fernandez},
  {Lenfestey}, {Li}, {Olguin}, {R{\"o}ck}, {de Villiers}, \&
  {Williams}}]{Peretto14}
{Peretto}, N., {Fuller}, G.~A., {Andr{\'e}}, P., {et~al.} 2014, \aap, 561, A83

\bibitem[{{Pillai} {et~al.}(2020){Pillai}, {Clemens}, {Reissl}, {Myers},
  {Kauffmann}, {Lopez-Rodriguez}, {Alves}, {Franco}, {Henshaw}, {Menten},
  {Nakamura}, {Seifried}, {Sugitani}, \& {Wiesemeyer}}]{Pillai20}
{Pillai}, T. G.~S., {Clemens}, D.~P., {Reissl}, S., {et~al.} 2020, Nature
  Astronomy, 4, 1195

\bibitem[{{Planck Collaboration Int. XXXII}(2016)}]{planck2016-a}
{Planck Collaboration Int. XXXII}. 2016, {A{\&}A}, 586, A135

\bibitem[{{Planck Collaboration Int. XXXIII}(2016)}]{planck2016-b}
{Planck Collaboration Int. XXXIII}. 2016, {A{\&}A}, 586, A136

\bibitem[{{Planck Collaboration Int.XXXIV}(2016)}]{planck2016-c}
{Planck Collaboration Int.XXXIV}. 2016, \aap, 586, A137

\bibitem[{{Pokhrel} {et~al.}(2016){Pokhrel}, {Gutermuth}, {Ali}, {Megeath},
  {Pipher}, {Myers}, {Fischer}, {Henning}, {Wolk}, {Allen}, \&
  {Tobin}}]{Pokhrel16}
{Pokhrel}, R., {Gutermuth}, R., {Ali}, B., {et~al.} 2016, \mnras, 461, 22

\bibitem[{{Pokhrel} {et~al.}(2020){Pokhrel}, {Gutermuth}, {Betti}, {Offner},
  {Myers}, {Megeath}, {Sokol}, {Ali}, {Allen}, {Allen}, {Dunham}, {Fischer},
  {Henning}, {Heyer}, {Hora}, {Pipher}, {Tobin}, \& {Wolk}}]{Pokhrel20}
{Pokhrel}, R., {Gutermuth}, R.~A., {Betti}, S.~K., {et~al.} 2020, \apj, 896, 60

\bibitem[{{Pokhrel} {et~al.}(2021){Pokhrel}, {Gutermuth}, {Krumholz},
  {Federrath}, {Heyer}, {Khullar}, {Megeath}, {Myers}, {Offner}, {Pipher},
  {Fischer}, {Henning}, \& {Hora}}]{Pokhrel21}
{Pokhrel}, R., {Gutermuth}, R.~A., {Krumholz}, M.~R., {et~al.} 2021, \apjl,
  912, L19

\bibitem[{{Ragan} {et~al.}(2014){Ragan}, {Henning}, {Tackenberg}, {Beuther},
  {Johnston}, {Kainulainen}, \& {Linz}}]{Ragan14}
{Ragan}, S.~E., {Henning}, T., {Tackenberg}, J., {et~al.} 2014, \aap, 568, A73

\bibitem[{{Rayner} {et~al.}(2017){Rayner}, {Griffin}, {Schneider}, {Motte},
  {K{\"o}nyves}, {Andr{\'e}}, {Di Francesco}, {Didelon}, {Pattle},
  {Ward-Thompson}, {Anderson}, {Benedettini}, {Bernard}, {Bontemps}, {Elia},
  {Fuente}, {Hennemann}, {Hill}, {Kirk}, {Marsh}, {Men'shchikov}, {Nguyen
  Luong}, {Peretto}, {Pezzuto}, {Rivera-Ingraham}, {Roy}, {Rygl},
  {S{\'a}nchez-Monge}, {Spinoglio}, {Tig{\'e}}, {Trevi{\~n}o-Morales}, \&
  {White}}]{Rayner2017}
{Rayner}, T.~S.~M., {Griffin}, M.~J., {Schneider}, N., {et~al.} 2017, \aap,
  607, A22

\bibitem[{{Roy} {et~al.}(2014){Roy}, {Andr{\'e}}, {Palmeirim}, {Attard},
  {K{\"o}nyves}, {Schneider}, {Peretto}, {Men'shchikov}, {Ward-Thompson},
  {Kirk}, {Griffin}, {Marsh}, {Abergel}, {Arzoumanian}, {Benedettini}, {Hill},
  {Motte}, {Nguyen Luong}, {Pezzuto}, {Rivera-Ingraham}, {Roussel}, {Rygl},
  {Spinoglio}, {Stamatellos}, \& {White}}]{Roy2014}
{Roy}, A., {Andr{\'e}}, P., {Palmeirim}, P., {et~al.} 2014, \aap, 562, A138

\bibitem[{{Santos} {et~al.}(2019){Santos}, {Chuss}, {Dowell}, {Houde},
  {Looney}, {Lopez Rodriguez}, {Novak}, {Ward-Thompson}, {Berthoud}, {Dale},
  {Guerra}, {Hamilton}, {Hanany}, {Harper}, {Henning}, {Jones}, {Lazarian},
  {Michail}, {Morris}, {Staguhn}, {Stephens}, {Tassis}, {Trinh}, {Van Camp},
  {Volpert}, \& {Wollack}}]{Fabio19}
{Santos}, F.~P., {Chuss}, D.~T., {Dowell}, C.~D., {et~al.} 2019, \apj, 882, 113

\bibitem[{{Schneider} {et~al.}(2012){Schneider}, {Csengeri}, {Hennemann},
  {Motte}, {Didelon}, {Federrath}, {Bontemps}, {Di Francesco}, {Arzoumanian},
  {Minier}, {Andr{\'e}}, {Hill}, {Zavagno}, {Nguyen-Luong}, {Attard},
  {Bernard}, {Elia}, {Fallscheer}, {Griffin}, {Kirk}, {Klessen}, {K{\"o}nyves},
  {Martin}, {Men'shchikov}, {Palmeirim}, {Peretto}, {Pestalozzi}, {Russeil},
  {Sadavoy}, {Sousbie}, {Testi}, {Tremblin}, {Ward-Thompson}, \&
  {White}}]{schneider2012}
{Schneider}, N., {Csengeri}, T., {Hennemann}, M., {et~al.} 2012, \aap, 540, L11

\bibitem[{{Schneider} \& {Elmegreen}(1979)}]{SchneiderElmegreen79}
{Schneider}, S. \& {Elmegreen}, B.~G. 1979, \apjs, 41, 87

\bibitem[{{Soler} {et~al.}(2016){Soler}, {Alves}, {Boulanger}, {Bracco},
  {Falgarone}, {Franco}, {Guillet}, {Hennebelle}, {Levrier}, {Martin}, \&
  {Miville-Desch{\^e}nes}}]{soler16}
{Soler}, J.~D., {Alves}, F., {Boulanger}, F., {et~al.} 2016, \aap, 596, A93

\bibitem[{{Tafalla} \& {Hacar}(2015)}]{TafallaHacar15}
{Tafalla}, M. \& {Hacar}, A. 2015, \aap, 574, A104

\bibitem[{{Trevi{\~n}o-Morales} {et~al.}(2019){Trevi{\~n}o-Morales}, {Fuente},
  {S{\'a}nchez-Monge}, {Kainulainen}, {Didelon}, {Suri}, {Schneider},
  {Ballesteros-Paredes}, {Lee}, {Hennebelle}, {Pilleri},
  {Gonz{\'a}lez-Garc{\'\i}a}, {Kramer}, {Garc{\'\i}a-Burillo}, {Luna},
  {Goicoechea}, {Tremblin}, \& {Geen}}]{TrevinoMorales2019}
{Trevi{\~n}o-Morales}, S.~P., {Fuente}, A., {S{\'a}nchez-Monge}, {\'A}.,
  {et~al.} 2019, \aap, 629, A81

\bibitem[{{V{\'a}zquez-Semadeni} {et~al.}(2017){V{\'a}zquez-Semadeni},
  {Gonz{\'a}lez-Samaniego}, \& {Col{\'\i}n}}]{Enrique2017}
{V{\'a}zquez-Semadeni}, E., {Gonz{\'a}lez-Samaniego}, A., \& {Col{\'\i}n}, P.
  2017, \mnras, 467, 1313

\bibitem[{{Ward-Thompson} {et~al.}(1994){Ward-Thompson}, {Scott}, {Hills}, \&
  {Andre}}]{Wardthompson94}
{Ward-Thompson}, D., {Scott}, P.~F., {Hills}, R.~E., \& {Andre}, P. 1994,
  \mnras, 268, 276

\bibitem[{{Williams} {et~al.}(2018){Williams}, {Peretto}, {Avison},
  {Duarte-Cabral}, \& {Fuller}}]{Williams18}
{Williams}, G.~M., {Peretto}, N., {Avison}, A., {Duarte-Cabral}, A., \&
  {Fuller}, G.~A. 2018, \aap, 613, A11

\bibitem[{{Zucker} {et~al.}(2018){Zucker}, {Battersby}, \&
  {Goodman}}]{Zucker2018}
{Zucker}, C., {Battersby}, C., \& {Goodman}, A. 2018, \apj, 864, 153

\end{thebibliography}

\begin{appendix}
\section{Structural decomposition maps}
\begin{figure}[ht]
\centering
\includegraphics[width=\linewidth]{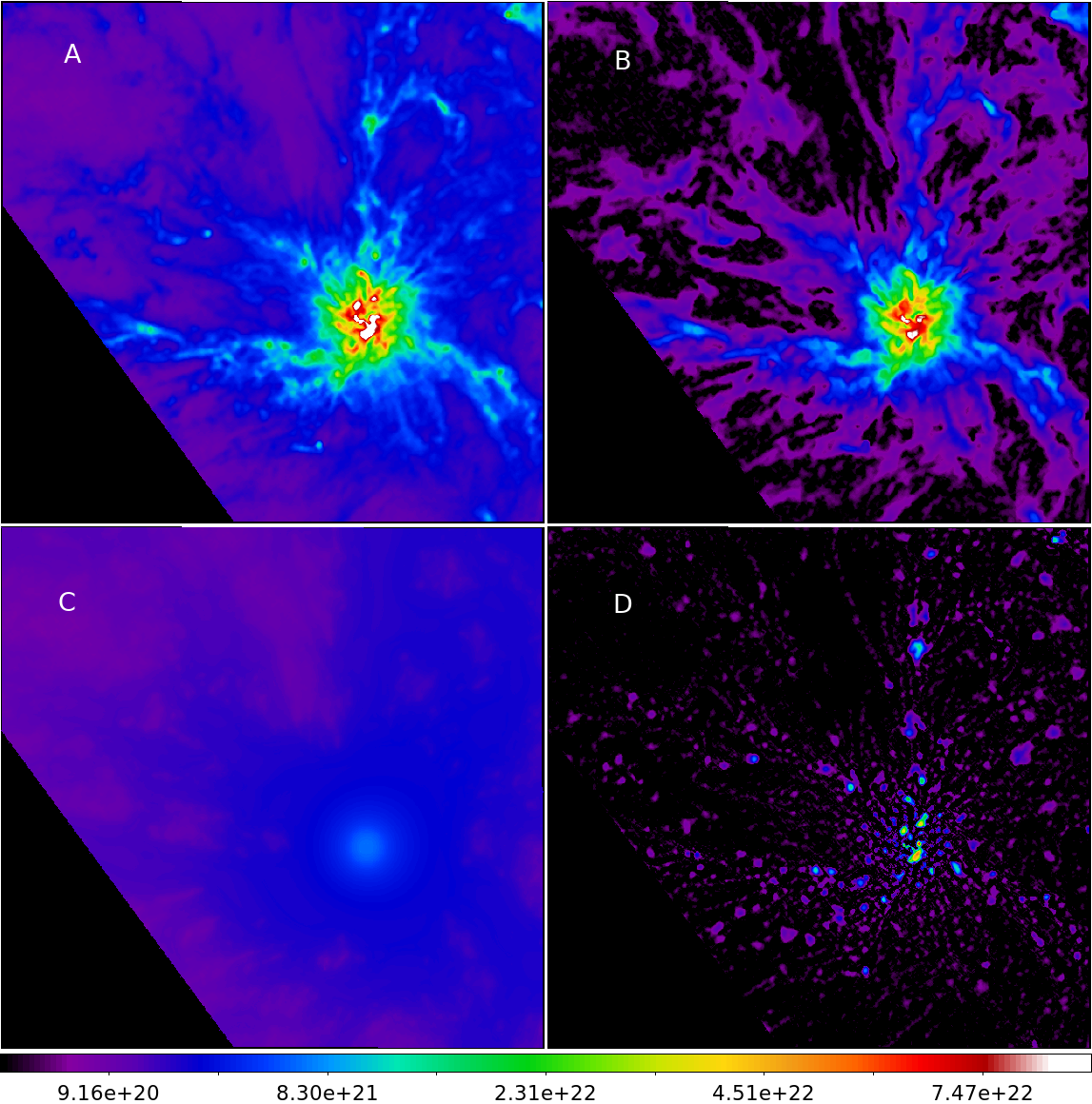}
\caption{Structural component decomposition. The images shown are: a) \herschel column density map at 18.2\arcsec, b) filaments, c) extended background, and  d) sources. The total flux in the image area is conserved with the column density map representing a sum of its structural components. The colour bar is common to all panels.}
\label{fig:struct}
\end{figure}

\begin{figure}[ht]
\centering
\includegraphics[width=\linewidth]{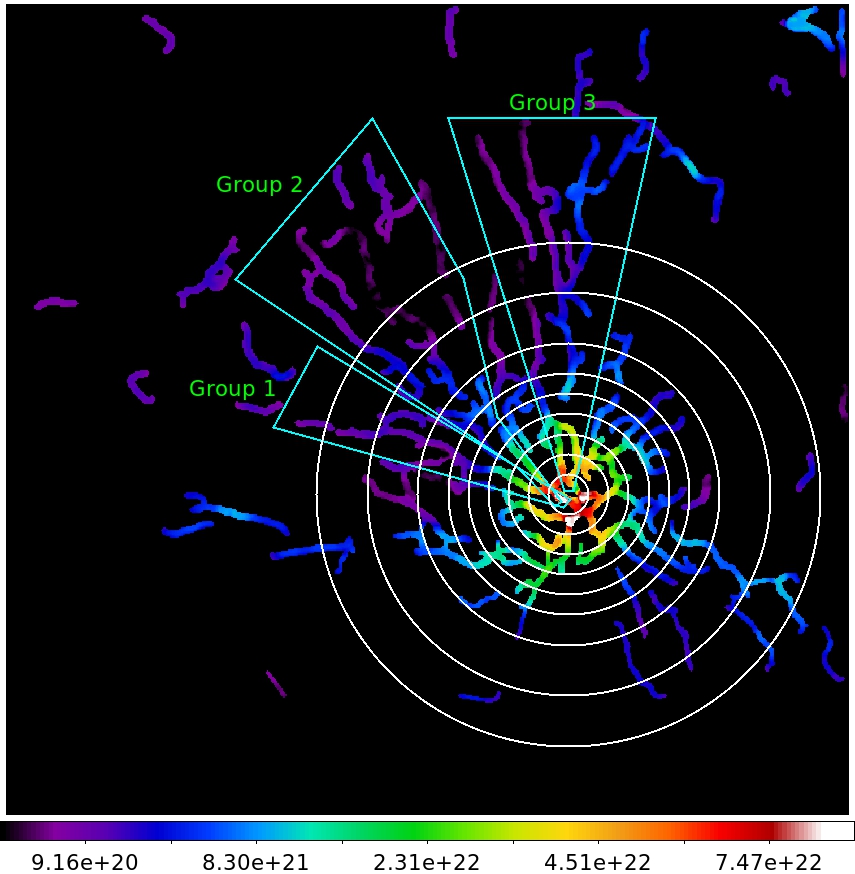}
\caption{Number of filaments that intersect with the white circular annuli drawn on the filament-component map masked by skeletons which were used to plot the black curve in Fig.\,5.}
\label{fig:circles}
\end{figure}

\begin{figure}[ht]
\centering
\includegraphics[width=\linewidth]{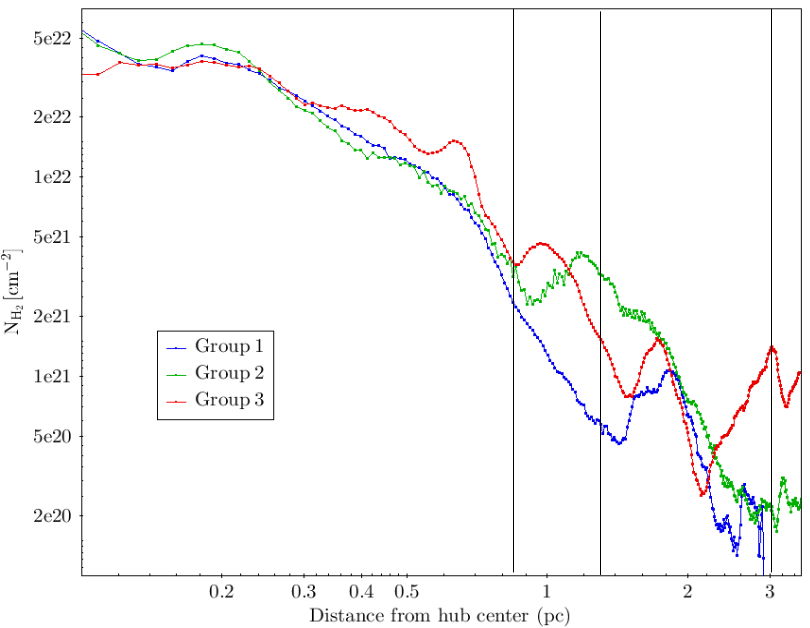}
\caption{Radial profiles of azimuthally averaged column density for the three groups of coalescing filaments identified in Fig.\,A.2. We note that these are profiles obtained on the filament-component map masked by detected skeletons. These groups encompass a larger number of filaments at larger distances while narrowing down on individual filaments closer to the centre. Readers should notice that the slopes of the radial profiles changes once at $\sim$0.8\,pc and again at $\sim$0.2\,pc. At larger distances from the centre, 
it is not possible to discern noticeable effects. The vertical lines at 3\,pc, 1.3\,pc, and 0.85\,pc correspond  to the two 
outermost red circles (nodes), the red circle just above the black cone, and the outer boundary of the black cone in Fig.\,2a, respectively. They demarcate the two steps of 
coalescence suggested in Sec.\,4.1}
\label{fig:N-ML}
\end{figure}

\end{appendix}

\end{document}